\documentclass[sigconf]{acmart}

\makeatletter
\def\@secfont{\sffamily\bfseries\Large}
\makeatother

\AtBeginDocument{%
  }

\usepackage[acronyms,nonumberlist,nopostdot,nomain,nogroupskip,acronymlists={hidden}]{glossaries}
\newglossary[algh]{hidden}{acrh}{acnh}{Hidden Acronyms}
\glsdisablehyper

\usepackage{tikz}
\usepackage{pgfplots}
\usepackage{glossaries}
\usepackage{soul}
\usepackage{xspace}
\usepackage{ragged2e}
\usepackage{subcaption}
\usetikzlibrary{patterns}
\usepackage{enumitem}

\newif\ifexttikz
\exttikzfalse

\ifexttikz
  \usetikzlibrary{external}
\fi

\usepackage{pgfplots}
\pgfplotsset{compat=newest}

\usepackage{enumitem}
\setlist{nosep} 

\usepackage{microtype}
\usepackage{balance} 

\newacronym{aoa}{AoA}{Angle of Arrival}
\newacronym{dapp}{dApp}{distributed Application}
\newacronym{cuphy}{cuPHY}{CUDA Physical layer}
\newacronym{cuda}{CUDA}{Compute Unified Device Architecture}
\newacronym{cubb}{cuBB}{CUDA Baseband}
\newacronym{onnx}{ONNX}{Open Neural Network Exchange}
\newacronym{ldpc}{LDPC}{Low Density Parity-Check Code}
\newacronym{crc}{CRC}{Cyclic Redundancy Check}
\newacronym{dnn}{DNN}{Deep Neural Network}
\newacronym{ort}{ORT}{ONNX Runtime}
\newacronym{tensorrt}{TensorRT}{Tensor RealTime}
\newacronym{cnn}{CNN}{Convolutional Neural Network}
\newacronym{cudnn}{cuDNN}{CUDA Deep Neural Network library}
\newacronym{cublas}{cuBLAS}{CUDA Basic Linear Algebra Subroutines}
\newacronym{hdf5}{HDF5}{Hierarchical Data Format version 5}
\newacronym{rmr}{RMR}{RIC Message Router}
\newacronym{3gpp}{3GPP}{3rd Generation Partnership Project}
\newacronym{4g}{4G}{4th generation}
\newacronym{5g}{5G}{fifth generation}
\newacronym{6g}{6G}{Sixth generation}
\newacronym{5gc}{5GC}{5G Core}
\newacronym{adc}{ADC}{Analog to Digital Converter}
\newacronym{aerpaw}{AERPAW}{Aerial Experimentation and Research Platform for Advanced Wireless}
\newacronym{ai}{AI}{Artificial Intelligence}
\newacronym{aimd}{AIMD}{Additive Increase Multiplicative Decrease}
\newacronym{am}{AM}{Acknowledged Mode}
\newacronym{amc}{AMC}{Adaptive Modulation and Coding}
\newacronym{amf}{AMF}{Access and Mobility Management Function}
\newacronym{ann}{ANN}{Artificial Neural Network}
\newacronym{aops}{AOPS}{Adaptive Order Prediction Scheduling}
\newacronym{api}{API}{Application Programming Interface}
\newacronym{apn}{APN}{Access Point Name}
\newacronym{aqm}{AQM}{Active Queue Management}
\newacronym{arc-ota}{ARC-OTA}{Aerial RAN CoLab Over-the-Air}
\newacronym{ausf}{AUSF}{Authentication Server Function}
\newacronym{avc}{AVC}{Advanced Video Coding}
\newacronym{awgn}{AWGN}{Additive White Gaussian Noise}
\newacronym{balia}{BALIA}{Balanced Link Adaptation Algorithm}
\newacronym{bbu}{BBU}{Base Band Unit}
\newacronym{bdp}{BDP}{Bandwidth-Delay Product}
\newacronym{ber}{BER}{Bit Error Rate}
\newacronym{bf}{BF}{Beamforming}
\newacronym{bler}{BLER}{Block Error Rate}
\newacronym{brr}{BRR}{Bayesian Ridge Regressor}
\newacronym{bsr}{BSR}{Buffer Status Report}
\newacronym{bs}{BS}{Base Station}
\newacronym{bpsk}{BPSK}{Binary Phase-shift keying}
\newacronym{bss}{BSS}{Business Support System}
\newacronym{ca}{CA}{Carrier Aggregation}
\newacronym{caas}{CaaS}{Connectivity-as-a-Service}
\newacronym{cb}{CB}{Code Block}
\newacronym{cc}{CC}{Congestion Control}
\newacronym{ccid}{CCID}{Congestion Control ID}
\newacronym{cco}{CC}{Carrier Component}
\newacronym{cdd}{CDD}{Cyclic Delay Diversity}
\newacronym{cdf}{CDF}{Cumulative Distribution Function}
\newacronym{cdn}{CDN}{Content Distribution Network}
\newacronym{cir}{CIR}{Channel Impulse Response}
\newacronym{cn}{CN}{Core Network}
\newacronym{codel}{CoDel}{Controlled Delay Management}
\newacronym{comac}{COMAC}{Converged Multi-Access and Core}
\newacronym{cord}{CORD}{Central Office Re-architected as a Datacenter}
\newacronym{cornet}{CORNET}{COgnitive Radio NETwork}
\newacronym{cosmos}{COSMOS}{Cloud Enhanced Open Software Defined Mobile Wireless Testbed for City-Scale Deployment}
\newacronym{cots}{COTS}{Commercial Off-the-Shelf}
\newacronym{cp}{CP}{Control Plane}
\newacronym{cpu}{CPU}{Central Processing Unit}
\newacronym{cqi}{CQI}{Channel Quality Information}
\newacronym{cr}{CR}{Cognitive Radio}
\newacronym{cran}{CRAN}{Cloud \gls{ran}}
\newacronym{crs}{CRS}{Cell Reference Signal}
\newacronym{csi}{CSI}{Channel State Information}
\newacronym{csirs}{CSI-RS}{Channel State Information - Reference Signal}
\newacronym{cu}{CU}{Central Unit}
\newacronym{d2tcp}{D$^2$TCP}{Deadline-aware Data center TCP}
\newacronym{d3}{D$^3$}{Deadline-Driven Delivery}
\newacronym{dac}{DAC}{Digital to Analog Converter}
\newacronym{dag}{DAG}{Directed Acyclic Graph}
\newacronym{darpa}{DARPA}{Defense Advanced Research Projects Agency}
\newacronym{das}{DAS}{Distributed Antenna System}
\newacronym{dash}{DASH}{Dynamic Adaptive Streaming over HTTP}
\newacronym{dc}{DC}{Dual Connectivity}
\newacronym{dccp}{DCCP}{Datagram Congestion Control Protocol}
\newacronym{dce}{DCE}{Direct Code Execution}
\newacronym{dci}{DCI}{Downlink Control Information}
\newacronym{dcl}{DCL}{Dear Colleague Letter}
\newacronym{dctcp}{DCTCP}{Data Center TCP}
\newacronym{dl}{DL}{Downlink}
\newacronym{dmr}{DMR}{Deadline Miss Ratio}
\newacronym{dmrs}{DMRS}{DeModulation Reference Signal}
\newacronym{drlcc}{DRL-CC}{Deep Reinforcement Learning Congestion Control}
\newacronym{drs}{DRS}{Discovery Reference Signal}
\newacronym{du}{DU}{Distributed Unit}
\newacronym{e2e}{E2E}{end-to-end}
\newacronym{e2sm}{E2SM}{E2 Service Model}
\newacronym{ecaas}{ECaaS}{Edge-Cloud-as-a-Service}
\newacronym{ecn}{ECN}{Explicit Congestion Notification}
\newacronym{edf}{EDF}{Earliest Deadline First}
\newacronym{eirp}{EIRP}{Effective Isotropic Radiated Power}
\newacronym{em}{EM}{Electro-Magnetic}
\newacronym{embb}{eMBB}{Enhanced Mobile Broadband}
\newacronym{empower}{EMPOWER}{EMpowering transatlantic PlatfOrms for advanced WirEless Research}
\newacronym{enb}{eNB}{evolved Node Base}
\newacronym{endc}{EN-DC}{E-UTRAN-\gls{nr} \gls{dc}}
\newacronym{epc}{EPC}{Evolved Packet Core}
\newacronym{eps}{EPS}{Evolved Packet System}
\newacronym{es}{ES}{Edge Server}
\newacronym{etsi}{ETSI}{European Telecommunications Standards Institute}
\newacronym[firstplural=Estimated Times of Arrival (ETAs)]{eta}{ETA}{Estimated Time of Arrival}
\newacronym{eutran}{E-UTRAN}{Evolved Universal Terrestrial Access Network}
\newacronym{faas}{FaaS}{Function-as-a-Service}
\newacronym{fapi}{FAPI}{Functional Application Platform Interface}
\newacronym{fcc}{FCC}{Federal Communications Commission}
\newacronym{fdd}{FDD}{Frequency Division Duplexing}
\newacronym{fdm}{FDM}{Frequency Division Multiplexing}
\newacronym{fdma}{FDMA}{Frequency Division Multiple Access}
\newacronym{fed4fire}{FED4FIRE+}{Federation 4 Future Internet Research and Experimentation Plus}
\newacronym{fir}{FIR}{Finite Impulse Response}
\newacronym{fit}{FIT}{Future \acrlong{iot}}
\newacronym{fpga}{FPGA}{Field Programmable Gate Array}
\newacronym{fr2}{FR2}{Frequency Range 2}
\newacronym{fs}{FS}{Fast Switching}
\newacronym{fscc}{FSCC}{Flow Sharing Congestion Control}
\newacronym{ftp}{FTP}{File Transfer Protocol}
\newacronym{fw}{FW}{Flow Window}
\newacronym{ga128}{Ga}{Golay Sequence type A}
\newacronym{ge}{GE}{Gaussian Elimination}
\newacronym{glfsr}{GLFSR}{Galois Linear Feedback Shift Register}
\newacronym{gnb}{gNB}{Next Generation Node Base}
\newacronym{gold}{Gold}{Gold}
\newacronym{gop}{GOP}{Group of Pictures}
\newacronym{gpr}{GPR}{Gaussian Process Regressor}
\newacronym{gpu}{GPU}{Graphics Processing Unit}
\newacronym{gtp}{GTP}{GPRS Tunneling Protocol}
\newacronym{gtpc}{GTP-C}{GPRS Tunnelling Protocol Control Plane}
\newacronym{gtpu}{GTP-U}{GPRS Tunnelling Protocol User Plane}
\newacronym{gtpv2c}{GTPv2-C}{\gls{gtp} v2 - Control}
\newacronym{gw}{GW}{Gateway}
\newacronym{harq}{HARQ}{Hybrid Automatic Repeat Request}
\newacronym{hetnet}{HetNet}{Heterogeneous Network}
\newacronym{hh}{HH}{Hard Handover}
\newacronym{hol}{HOL}{Head-of-Line}
\newacronym{hqf}{HQF}{Highest-quality-first}
\newacronym{hss}{HSS}{Home Subscription Server}
\newacronym{http}{HTTP}{HyperText Transfer Protocol}
\newacronym{ia}{IA}{Initial Access}
\newacronym{iab}{IAB}{Integrated Access and Backhaul}
\newacronym{ic}{IC}{Incident Command}
\newacronym{ietf}{IETF}{Internet Engineering Task Force}
\newacronym{ifw}{IFW}{Interference Free Window}
\newacronym{imsi}{IMSI}{International Mobile Subscriber Identity}
\newacronym{imt}{IMT}{International Mobile Telecommunication}
\newacronym{iot}{IoT}{Internet of Things}
\newacronym{ip}{IP}{Internet Protocol}
\newacronym{iq}{I/Q}{In-phase and Quadrature}
\newacronym{itu}{ITU}{International Telecommunication Union}
\newacronym{iw}{IW}{Interference Whitening}
\newacronym{kpi}{KPI}{Key Performance Indicator}
\newacronym{kpm}{KPM}{Key Performance Measurement}
\newacronym{kvm}{KVM}{Kernel-based Virtual Machine}
\newacronym{leo}{LEO}{Low Earth Orbit}
\newacronym{los}{LOS}{Line-of-Sight}
\newacronym{ls}{LS}{Least Squares}
\newacronym{lsm}{LSM}{Link-to-System Mapping}
\newacronym{lstm}{LSTM}{Long Short Term Memory}
\newacronym{lte}{LTE}{Long Term Evolution}
\newacronym{lxc}{LXC}{Linux Container}
\newacronym{m2m}{M2M}{Machine to Machine}
\newacronym{mac}{MAC}{Medium Access Control}
\newacronym{manet}{MANET}{Mobile Ad Hoc Network}
\newacronym{mano}{MANO}{Management and Orchestration}
\newacronym{mc}{MC}{Multi-Connectivity}
\newacronym{mcc}{MCC}{Mobile Cloud Computing}
\newacronym{mchem}{MCHEM}{Massive Channel Emulator}
\newacronym{mcs}{MCS}{Modulation and Coding Scheme}
\newacronym{mec}{MEC}{Multi-access Edge Computing}
\newacronym{mec2}{MEC}{Mobile Edge Cloud}
\newacronym{mfc}{MFC}{Mobile Fog Computing}
\newacronym{mi}{MI}{Mutual Information}
\newacronym{mib}{MIB}{Master Information Block}
\newacronym{miesm}{MIESM}{Mutual Information Based Effective SINR}
\newacronym{mimo}{MIMO}{Multiple Input, Multiple Output}
\newacronym{mgen}{MGEN}{Multi-Generator}
\newacronym{ml}{ML}{Machine Learning}
\newacronym{mlr}{MLR}{Maximum-local-rate}
\newacronym[plural=\gls{mme}s,firstplural=Mobility Management Entities (MMEs)]{mme}{MME}{Mobility Management Entity}
\newacronym{mmtc}{mMTC}{Massive Machine-Type Communications}
\newacronym{mmwave}{mmWave}{millimeter wave}
\newacronym{mpdccp}{MP-DCCP}{Multipath Datagram Congestion Control Protocol}
\newacronym{mptcp}{MPTCP}{Multipath TCP}
\newacronym{mr}{MR}{Maximum Rate}
\newacronym{mrdc}{MR-DC}{Multi \gls{rat} \gls{dc}}
\newacronym{mse}{MSE}{Mean Square Error}
\newacronym{mss}{MSS}{Maximum Segment Size}
\newacronym{mt}{MT}{Mobile Termination}
\newacronym{mtd}{MTD}{Machine-Type Device}
\newacronym{mtu}{MTU}{Maximum Transmission Unit}
\newacronym{mumimo}{MU-MIMO}{Multi-user \gls{mimo}}
\newacronym{mvno}{MVNO}{Mobile Virtual Network Operator}
\newacronym{nalu}{NALU}{Network Abstraction Layer Unit}
\newacronym{nas}{NAS}{Network Attached Storage}
\newacronym{nbiot}{NB-IoT}{Narrow Band IoT}
\newacronym{nfv}{NFV}{Network Function Virtualization}
\newacronym{nfvi}{NFVI}{Network Function Virtualization Infrastructure}
\newacronym{nic}{NIC}{Network Interface Card}
\newacronym{nlos}{NLOS}{Non-Line-of-Sight}
\newacronym{now}{NOW}{Non Overlapping Window}
\newacronym{nrdz}{NRDZ}{National Radio Dynamic Zone}
\newacronym{nsf}{NSF}{National Science Foundation}
\newacronym{nsm}{NSM}{Network Service Mesh}
\newacronym{nr}{NR}{New Radio}
\newacronym{nrf}{NRF}{Network Repository Function}
\newacronym{nsa}{NSA}{Non Stand Alone}
\newacronym{nse}{NSE}{Network Slicing Engine}
\newacronym{nssf}{NSSF}{Network Slice Selection Function}
\newacronym{ntp}{NTP}{Network Time Protocol}
\newacronym{o2i}{O2I}{Outdoor to Indoor}
\newacronym{oai}{OAI}{OpenAirInterface}
\newacronym{oaic}{OAIC}{Open AI Cellular}
\newacronym{oaicn}{OAI-CN}{\gls{oai} \acrlong{cn}}
\newacronym{oairan}{OAI-RAN}{\acrlong{oai} \acrlong{ran}}
\newacronym{oam}{OAM}{Operations, Administration and Maintenance}
\newacronym[plural=\gls{obu}s,firstplural=Onboard Units (OBUs)]{obu}{OBU}{Onboard Unit}
\newacronym{ofdm}{OFDM}{Orthogonal Frequency Division Multiplexing}
\newacronym{olia}{OLIA}{Opportunistic Linked Increase Algorithm}
\newacronym{omec}{OMEC}{Open Mobile Evolved Core}
\newacronym{onap}{ONAP}{Open Network Automation Platform}
\newacronym{onf}{ONF}{Open Networking Foundation}
\newacronym{onos}{ONOS}{Open Networking Operating System}
\newacronym{oom}{OOM}{\gls{onap} Operations Manager}
\newacronym{opnfv}{OPNFV}{Open Platform for \gls{nfv}}
\newacronym{orbit}{ORBIT}{Open-Access Research Testbed for Next-Generation Wireless Networks}
\newacronym{os}{OS}{Operating System}
\newacronym{osc}{OSC}{O-RAN Software Community}
\newacronym{osm}{OSM}{Open Street Map}
\newacronym{oss}{OSS}{Operations Support System}
\newacronym{pa}{PA}{Position-aware}
\newacronym{pase}{PASE}{Prioritization, Arbitration, and Self-adjusting Endpoints}
\newacronym{pawr}{PAWR}{Platforms for Advanced Wireless Research}
\newacronym{pbch}{PBCH}{Physical Broadcast Channel}
\newacronym{pci}{PCI}{Peripheral Component Interconnect}
\newacronym{pcef}{PCEF}{Policy and Charging Enforcement Function}
\newacronym{pcfich}{PCFICH}{Physical Control Format Indicator Channel}
\newacronym{pcrf}{PCRF}{Policy and Charging Rules Function}
\newacronym{pdcch}{PDCCH}{Physical Downlink Control Channel}
\newacronym{pdcp}{PDCP}{Packet Data Convergence Protocol}
\newacronym{pdsch}{PDSCH}{Physical Downlink Shared Channel}
\newacronym{pdu}{PDU}{Packet Data Unit}
\newacronym{pdp}{PDP}{Power Delay Profile}
\newacronym{pf}{PF}{Proportional Fair}
\newacronym{pgw}{PGW}{Packet Gateway}
\newacronym{ph}{PH}{Power Headroom}
\newacronym{phich}{PHICH}{Physical Hybrid ARQ Indicator Channel}
\newacronym{phy}{PHY}{Physical}
\newacronym{pl}{PL}{Path Loss}
\newacronym{pmch}{PMCH}{Physical Multicast Channel}
\newacronym{pmi}{PMI}{Precoding Matrix Indicators}
\newacronym{powder}{POWDER}{Platform for Open Wireless Data-driven Experimental Research}
\newacronym{ppo}{PPO}{Proximal Policy Optimization}
\newacronym{ppp}{PPP}{Poisson Point Process}
\newacronym{prach}{PRACH}{Physical Random Access Channel}
\newacronym{prb}{PRB}{Physical Resource Block}
\newacronym{psnr}{PSNR}{Peak Signal to Noise Ratio}
\newacronym{pss}{PSS}{Primary Synchronization Signal}
\newacronym{pucch}{PUCCH}{Physical Uplink Control Channel}
\newacronym{pusch}{PUSCH}{Physical Uplink Shared Channel}
\newacronym{qam}{QAM}{Quadrature Amplitude Modulation}
\newacronym{qci}{QCI}{\gls{qos} Class Identifier}
\newacronym{qoe}{QoE}{Quality of Experience}
\newacronym{qos}{QoS}{Quality of Service}
\newacronym{qtgui}{QT-GUI}{QT Graphical User Interface}
\newacronym{qsfp28}{QSFP28}{Quad Small Form-factor Pluggable 28}
\newacronym{quic}{QUIC}{Quick UDP Internet Connections}
\newacronym{rach}{RACH}{Random Access Channel}
\newacronym{ran}{RAN}{Radio Access Network}
\newacronym[firstplural=Radio Access Technologies (RATs)]{rat}{RAT}{Radio Access Technology}
\newacronym{rcn}{RCN}{Research Coordination Network}
\newacronym{rdma}{RDMA}{Remote Direct Memory Access}
\newacronym{rec}{REC}{Radio Edge Cloud}
\newacronym{red}{RED}{Random Early Detection}
\newacronym{renew}{RENEW}{Reconfigurable Eco-system for Next-generation End-to-end Wireless}
\newacronym{rf}{RF}{Radio Frequency}
\newacronym{rfc}{RFC}{Request for Comments}
\newacronym{rfr}{RFR}{Random Forest Regressor}
\newacronym{ric}{RIC}{RAN Intelligent Controller}
\newacronym{near-rt-ric}{near-RT-RIC}{near-RT-\gls{ric}}
\newacronym{non-rt}{non-RT}{non-Real-Time}
\newacronym{near-rt}{near-RT}{near-Real-Time}
\newacronym{rlc}{RLC}{Radio Link Control}
\newacronym{rlf}{RLF}{Radio Link Failure}
\newacronym{rlnc}{RLNC}{Random Linear Network Coding}
\newacronym{rmse}{RMSE}{Root Mean Squared Error}
\newacronym{rnis}{RNIS}{Radio Network Information Service}
\newacronym{rnn}{RNN}{Recurrent Neural Network}
\newacronym{rr}{RR}{Round Robin}
\newacronym{rrc}{RRC}{Radio Resource Control}
\newacronym{rrm}{RRM}{Radio Resource Management}
\newacronym{rru}{RRU}{Remote Radio Unit}
\newacronym{rs}{RS}{Remote Server}
\newacronym{rsrp}{RSRP}{Reference Signal Received Power}
\newacronym{rsrq}{RSRQ}{Reference Signal Received Quality}
\newacronym{rss}{RSS}{Received Signal Strength}
\newacronym{rssi}{RSSI}{Received Signal Strength Indicator}
\newacronym{rsu}{RSU}{Road-Side Unit}
\newacronym{rtt}{RTT}{Round Trip Time}
\newacronym{ru}{RU}{Radio Unit}
\newacronym{rw}{RW}{Receive Window}
\newacronym{rx}{RX}{Receiver}
\newacronym{s1ap}{S1AP}{S1 Application Protocol}
\newacronym{sa}{SA}{standalone}
\newacronym{sack}{SACK}{Selective Acknowledgment}
\newacronym{sap}{SAP}{Service Access Point}
\newacronym{sas}{SAS}{Spectrum Access System}
\newacronym{sc2}{SC2}{Spectrum Collaboration Challenge}
\newacronym{scef}{SCEF}{Service Capability Exposure Function}
\newacronym{sch}{SCH}{Secondary Cell Handover}
\newacronym{scoot}{SCOOT}{Split Cycle Offset Optimization Technique}
\newacronym{sfp+}{SFP+}{Small Form-factor Pluggable Plus}
\newacronym{sctp}{SCTP}{Stream Control Transmission Protocol}
\newacronym{sdap}{SDAP}{Service Data Adaptation Protocol}
\newacronym{sd}{SD}{Standard Deviation}
\newacronym{sdk}{SDK}{Software Development Kit}
\newacronym{sdm}{SDM}{Space Division Multiplexing}
\newacronym{sdma}{SDMA}{Spatial Division Multiple Access}
\newacronym{sdn}{SDN}{Software-defined Networking}
\newacronym{sdr}{SDR}{Software-defined Radio}
\newacronym{seba}{SEBA}{SDN-Enabled Broadband Access}
\newacronym{sgsn}{SGSN}{Serving GPRS Support Node}
\newacronym{sgw}{SGW}{Service Gateway}
\newacronym{si}{SI}{Study Item}
\newacronym{sib}{SIB}{Secondary Information Block}
\newacronym{sinr}{SINR}{Signal to Interference plus Noise Ratio}
\newacronym{sip}{SIP}{Session Initiation Protocol}
\newacronym{siso}{SISO}{Single Input, Single Output}
\newacronym{sla}{SLA}{Service Level Agreement}
\newacronym{sm}{SM}{Saturation Mode}
\newacronym{smf}{SMF}{Session Management Function}
\newacronym{smo}{SMO}{Service Management and Orchestration}
\newacronym{sms}{SMS}{Short Message Service}
\newacronym{smsgmsc}{SMS-GMSC}{\gls{sms}-Gateway}
\newacronym{snr}{SNR}{Signal-to-Noise-Ratio}
\newacronym{son}{SON}{Self-Organizing Network}
\newacronym{sptcp}{SPTCP}{Single Path TCP}
\newacronym{srb}{SRB}{Service Radio Bearer}
\newacronym{srn}{SRN}{Standard Radio Node}
\newacronym{srs}{SRS}{Sounding Reference Signal}
\newacronym{ss}{SS}{Synchronization Signal}
\newacronym{sss}{SSS}{Secondary Synchronization Signal}
\newacronym{st}{ST}{Spanning Tree}
\newacronym{svc}{SVC}{Scalable Video Coding}
\newacronym{synce}{SyncE}{Synchronous Ethernet}
\newacronym{tb}{TB}{Transport Block}
\newacronym{tcp}{TCP}{Transmission Control Protocol}
\newacronym{tdd}{TDD}{Time Division Duplexing}
\newacronym{tdm}{TDM}{Time Division Multiplexing}
\newacronym{tdma}{TDMA}{Time Division Multiple Access}
\newacronym{tfl}{TfL}{Transport for London}
\newacronym{tfrc}{TFRC}{TCP-Friendly Rate Control}
\newacronym{tft}{TFT}{Traffic Flow Template}
\newacronym{tgen}{TGEN}{Traffic Generator}
\newacronym{tip}{TIP}{Telecom Infra Project}
\newacronym{tm}{TM}{Transparent Mode}
\newacronym{to}{TO}{Telco Operator}
\newacronym{toa}{ToA}{Time of Arrival}
\newacronym{tl}{TL}{Transfer Learning}
\newacronym{tr}{TR}{Technical Report}
\newacronym{trp}{TRP}{Transmitter Receiver Pair}
\newacronym{ts}{TS}{Technical Specification}
\newacronym{tti}{TTI}{Transmission Time Interval}
\newacronym{ttt}{TTT}{Time-to-Trigger}
\newacronym{tx}{TX}{Transmitter}
\newacronym{uas}{UAS}{Unmanned Aerial System}
\newacronym{uav}{UAV}{Unmanned Aerial Vehicle}
\newacronym{udm}{UDM}{Unified Data Management}
\newacronym{udp}{UDP}{User Datagram Protocol}
\newacronym{udr}{UDR}{Unified Data Repository}
\newacronym{ue}{UE}{User Equipment}
\newacronym{uhd}{UHD}{\gls{usrp} Hardware Driver}
\newacronym{ul}{UL}{uplink}
\newacronym{um}{UM}{Unacknowledged Mode}
\newacronym{uml}{UML}{Unified Modeling Language}
\newacronym{upa}{UPA}{Uniform Planar Array}
\newacronym{upf}{UPF}{User Plane Function}
\newacronym{urllc}{URLLC}{Ultra Reliable and Low Latency Communication}
\newacronym{usa}{U.S.}{United States}
\newacronym{usim}{USIM}{Universal Subscriber Identity Module}
\newacronym{usrp}{USRP}{Universal Software Radio Peripheral}
\newacronym{utc}{UTC}{Urban Traffic Control}
\newacronym{vim}{VIM}{Virtualization Infrastructure Manager}
\newacronym{vlan}{VLAN}{Virtual Local Area Network}
\newacronym{vm}{VM}{Virtual Machine}
\newacronym{vnf}{VNF}{Virtual Network Function}
\newacronym{volte}{VoLTE}{Voice over \gls{lte}}
\newacronym{voltha}{VOLTHA}{Virtual OLT HArdware Abstraction}
\newacronym{vr}{VR}{Virtual Reality}
\newacronym{vran}{vRAN}{Virtualized \gls{ran}}
\newacronym{vss}{VSS}{Video Streaming Server}
\newacronym{wbf}{WBF}{Wired Bias Function}
\newacronym{wf}{WF}{Wired-first}
\newacronym{wi}{WI}{Wireless InSite}
\newacronym{wlan}{WLAN}{Wireless Local Area Network}
\newacronym{pnf}{PNF}{Physical Network Function}
\newacronym{drl}{DRL}{Deep Reinforcement Learning}
\newacronym{mtc}{MTC}{Machine-type Communications}
\newacronym{v2x}{V2X}{Vehicle-to-everything}
\newacronym{cast}{\textit{CaST}}{Channel emulation generator and Sounder Toolchain}
\newacronym{abr}{ABR}{Adaptive Bitrate Streaming}
\newacronym{arc}{ARC}{Aerial RAN CoLab}
\newacronym{dsp}{DSP}{Digital Signal Processing}
\newacronym{ota}{OTA}{Over-the-Air}
\newacronym{bom}{BoM}{Bill of Materials}
\newacronym{frand}{FRAND}{Fair, Reasonable, And Non-Discriminatory}
\newacronym{nvipc}{NVIPC}{NVIDIA Inter-Process Communication}
\newacronym{hdr}{HDR}{High Dynamic Range}
\newacronym{ipc}{IPC}{Inter-Process Communication}
\newacronym{uci}{UCI}{Uplink Control Indication}
\newacronym{cbrs}{CBRS}{Citizen Broadband Radio Service}
\newacronym{ptp}{PTP}{Precision Timing Protocol}
\newacronym{scf}{SCF}{Small Cell Forum}
\newacronym{re}{RE}{Resource Element}
\newacronym{fp16}{FP16}{Float16}
\newacronym{fp32}{FP32}{Float32}
\newacronym{int32}{INT32}{32-bit Integer}
\newacronym{tdl}{TDL}{Tapped Delay Line}
\newacronym{fh}{FH}{Front-haul}
\newacronym{ta}{TA}{Timing Advance}
\newacronym{cfo}{CFO}{Carrier Frequency Offset}
\newacronym{sir}{SIR}{Signal to Interference Ratio}
\newacronym{mmse}{MMSE}{Minimum Mean Square Error}
\newacronym{irc}{IRC}{Interference Rejection Combining}
\newacronym{tf}{TF}{Tensorflow}
\newacronym{ndi}{NDI}{New Data Indicator}
\newacronym{isac}{ISAC}{Integrated Sensing and Communication}
\newacronym{moe}{MoE}{Mixture-of-Experts}
\acmDOI{XXXXXXX.XXXXXXX}

\acmConference[MobiHoc '26]{MobiHoc '26}{Nov 23--26, 2026}{Tokyo, Japan}

\acmISBN{978-1-4503-XXXX-X/18/06}

\newcommand{\switch}{ARCHES\xspace}

\usepackage{tikzpagenodes,etoolbox}
\usetikzlibrary{calc}
\usepackage[contents={}]{background}

\usepackage{eso-pic}

\pgfplotsset{compat=1.18} 
\begin{document}

\AddToShipoutPictureBG*{%
  \AtPageUpperLeft{%
    \raisebox{-2cm}{%
      \hspace*{\dimexpr(\paperwidth-0.9\paperwidth)/2\relax}%
      \parbox{0.9\paperwidth}{%
        \centering\footnotesize
        This work has been submitted to the ACM for possible publication.
        Copyright may be transferred without notice, after which this version may no longer be accessible.
      }%
    }%
  }%
}

\title{\switch: Adaptive Real-Time Switching of AI Models for the RAN}

\author[N. Neasamoni Santhi, D. Villa, M. Polese, S. D'Oro, Y. Lee, K. Furueda, T. Melodia]{Neagin Neasamoni Santhi$^{\star}$, Davide Villa$^{\star}$$^{\ddagger}$, Michele Polese$^{\star}$, Salvatore D'Oro$^{\star}$, \\Yunseong Lee$^{\dagger}$, Koichiro Furueda$^{\dagger}$, Tommaso Melodia$^{\star}$}
\affiliation{%
  \country{$^{\star}$Institute for Intelligent Network Systems, Northeastern University, Boston, MA, U.S.A. \\ \{neasamonisanthi.n, villa.d, m.polese, s.doro, t.melodia\}@northeastern.edu}\\
  \institution{$^{\dagger}$Technology Unit, SoftBank Corp., Tokyo, Japan\\ {yunseong.lee.1992@ieee.org}, {koichiro.furueda@g.softbank.co.jp}}
  \city{$^{\ddagger}$NVIDIA Corporation, Santa Clara, CA, U.S.A.\\ 
  {dvilla@nvidia.com}}\\
}

\begin{abstract}

\gls{ai} has become a powerful tool for model-free \gls{ran} signal processing and optimization. However, it is challenging to design and train a single model that generalizes reliably across all radio environments. Specialized AI models often outperform conventional algorithms only under specific channel conditions, while their substantially higher compute and energy cost makes unconditional execution impractical at the base station. This creates a need for real-time expert switching: dynamically activating the most appropriate AI or conventional signal-processing expert based on current network conditions.

To address this, we propose \switch (\textbf{A}daptive 
\textbf{R}eal-time \textbf{C}UDA \textbf{H}ot-swapping of 
\textbf{E}xperts in the \gls{ran} \textbf{S}tack), a framework that 
hosts multiple \gls{ai}-based and conventional signal processing experts within a \gls{gpu}-accelerated \gls{phy} pipeline and 
dynamically selects the most appropriate expert at slot-boundary 
granularity, without dropping or corrupting in-flight data. 
\switch includes a lightweight \gls{cuda} switch kernel for zero-gap 
output selection, a dApp-based control plane that collects 
cross-layer telemetry and drives the switching policy, and a reusable process for telemetry selection and policy 
design based on controlled perturbation, monotonicity filtering, 
and hierarchical clustering.

We validate \switch on \gls{ul} channel estimation, switching 
between an \gls{ai}-based and a \gls{mmse} estimator in response to 
changing propagation and interference conditions. The prototype, 
implemented on the X5G platform with NVIDIA Aerial and \gls{oai}, demonstrates median \gls{ul} \gls{phy} 
throughput gains of 5.32\% and 7.23\% under good and poor 
conditions, respectively, with an end-to-end control-loop 
latency of $\approx$140~$\mu$s and sub-microsecond decision 
inference. Under good conditions, defaulting to \gls{mmse} saves 
15.8~W of GPU power (9.6\%) and 17 percentage points of GPU 
utilization versus unconditional \gls{ai} execution, validating the 
performance-per-watt tradeoff that motivates adaptive expert selection.

\end{abstract}


\newlength\fwidth
\newlength\fheight

\glsresetall
\glsunset{nr}

\makeatletter
\def\@ACM@copyright@check@cc{}
\makeatother

\maketitle

\glsresetall
\glsunset{nr}
\glsunset{5g}
\glsunset{dapp}

\vspace{-.24cm}
\section{Introduction}
\label{sec:intro}

\gls{ai}-based techniques have demonstrated significant potential 
across a range of \gls{phy} and \gls{ran} functions in 5G and 
beyond~\cite{3gpp38843,chen2021ai}. In channel estimation, 
neural-network-based approaches can capture complex propagation 
characteristics that traditional linear estimators struggle to 
model~\cite{soltani2019channelnet,feriani2023cebed}. In beam 
management, \gls{ai} enables faster and more accurate beam prediction, 
reducing overhead from exhaustive beam 
sweeps~\cite{wang2019mmwave}. In positioning, learning-based 
methods improve localization accuracy by exploiting spatial 
features of the radio environment beyond conventional 
geometry-based approaches~\cite{wang2017csi}. In resource 
allocation and scheduling, \gls{ai}-driven strategies can adapt to 
dynamic traffic patterns and interference conditions more 
effectively than rule-based heuristics~\cite{bonati2021intelligence}.

A common requirement across all of these functions is real-time processing under strict latency constraints, as they operate within the tight timing budgets of the \gls{phy} and \gls{mac} layers. Such constraints demand lightweight models capable of fast inference, but their limited capacity often prevents them from generalizing across diverse channel conditions, spanning indoor and outdoor environments, pedestrian versus vehicular mobility, rural and urban deployments, and small and large cells. As a result, they tend to be specialized to a subset of conditions, where they can outperform traditional approaches, but may fail when operating outside their area of expertise~\cite{corgan2024site,verdecia2024enhancing}.

Crucially, even when an \gls{ai} model outperforms its conventional counterpart across all observed conditions, operational considerations may still favor selective activation. Our measurements on a production \gls{gpu}-accelerated \gls{gnb} show that an \gls{ai}-based channel estimator draws 15.8~W more GPU power and consumes 17 percentage points more GPU utilization than its conventional counterpart under favorable conditions, for a throughput gain of only $\sim$5\%. At scale, with dozens of cells and multiple \glspl{ue} per cell, unconditional \gls{ai} execution becomes impractical: the compute and energy cost per marginal bit of throughput does not justify always-on inference. As \gls{ai} components proliferate across the RAN, this performance-per-watt tradeoff will force operators to selectively activate \gls{ai} experts only when the performance benefit warrants the additional resource expenditure.

A common approach to address this lack of generalization is the \gls{moe} paradigm~\cite{jacobs1991adaptive, shazeer2017outrageously}, in which multiple specialized models are maintained simultaneously and incoming inputs are routed to the most suitable expert based on current channel conditions.  While classical MoE routes inputs among sub-networks within a single model, in this paper we extend the concept to switching between independent signal processing modules in a live pipeline.

However, applying this idea to real-time signal processing functions in the RAN introduces additional challenges. The routing decision itself must be made under strict latency constraints, requiring the system to detect changing conditions and switch between experts at \gls{phy}-timescale using inputs from different parts of the protocol stack and processing pipeline. Beyond decision latency, switching must occur seamlessly within a live processing pipeline without dropping or corrupting in-flight data, since each \gls{ul} slot carries irreplaceable user traffic that cannot be buffered or replayed. Furthermore, the decision must be informed by cross-layer metrics spanning \gls{phy}, \gls{mac}, and higher layers, requiring tightly coordinated telemetry collection across disaggregated RAN components operating on different timescales.

Motivated by this, we propose \textbf{\switch} (\textbf{A}daptive \textbf{R}eal-time \textbf{C}UDA \textbf{H}ot-swapping of \textbf{E}xperts in the RAN \textbf{S}tack), a general-purpose framework for real-time expert switching inside GPU-accelerated 5G \gls{phy} pipelines. \switch enables the simultaneous operation of multiple \gls{ai}-based and conventional signal processing experts and dynamically selects the most appropriate one based on current network conditions, without disrupting data flow or introducing processing gaps.
The framework continuously monitors live RAN metrics through a dApp-based control plane, evaluates prevailing network conditions using a lightweight policy, and selects the most suitable expert at slot boundaries.


\begin{figure}[htbp]
  \centering
      \includegraphics[width=1\columnwidth]{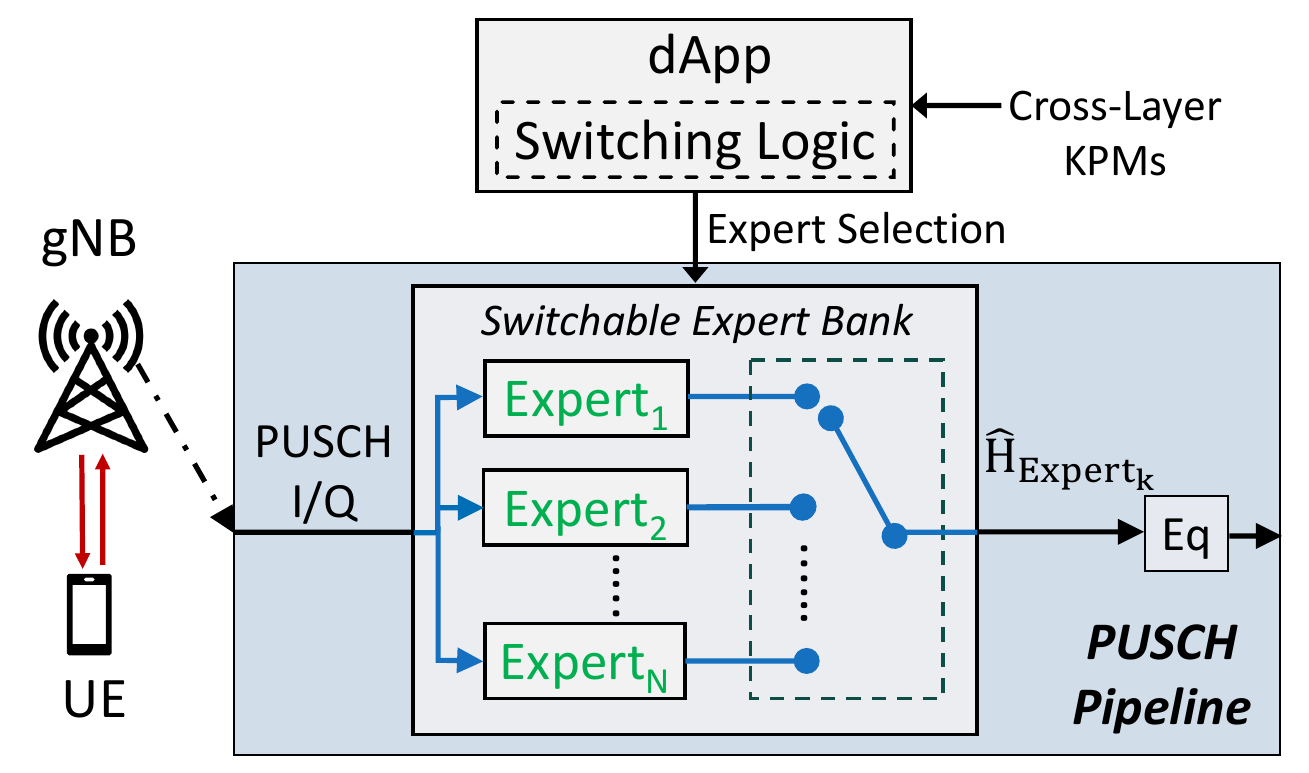}
  \caption{High-level system architecture of \switch. Multiple signal processing experts execute within the \gls{phy} pipeline. A dApp collects cross-layer KPMs and drives the switching logic, which selects the appropriate expert output. Shown here for channel estimation; the same architecture applies to other PHY functions.}
  \label{fig:sys_archi}
\end{figure}

As a first instantiation, we validate \switch on \gls{ul} channel estimation, where the framework switches between an \gls{ai}-based and a \gls{mmse} estimator in response to changing propagation and interference conditions. Channel estimation serves only as the first validating case study; the core contribution is the switching mechanism itself. The expert bank, the \gls{cuda} switch kernel, the dApp-based control path, and the telemetry-driven policy design are designed to generalize to any \gls{phy}/\gls{mac} function where multiple processing strategies coexist and the optimal choice depends on current network conditions.

Our main contributions are as follows:
\begin{itemize}
    \item We establish the design principles for real-time expert switching in GPU-accelerated \gls{phy} pipelines, including dual execution modes balancing switching latency and resource efficiency, memory aliasing for a uniform downstream interface, slot-boundary timing semantics, and fail-safe defaults.
    \item We implement the \switch framework: a generic CUDA switch kernel and dApp-based control path with structured host-to-device propagation and E3-based telemetry collection.
    \item We present a reusable process for selecting telemetry and switching policy design, based on controlled perturbation, monotonicity filtering, and redundancy reduction via hierarchical clustering.
  \item We validate \switch through a case-study instantiation on \gls{ul} channel estimation, demonstrating sub-microsecond decision latency, an end-to-end control loop of $\sim$140 $\mu$s, throughput gains, and quantified \gls{gpu} power and utilization savings via extensive \gls{ota} experiments on the X5G platform.
\end{itemize}

The remainder of the paper is organized as follows. Section~\ref{sec:principles} presents the design principles for real-time expert switching. Section~\ref{sec:framework} details the \switch framework. Section~\ref{sec:methodology} describes the methodology for policy design. Section~\ref{sec:casestudy} presents the channel estimation case study. Section~\ref{sec:eval} describes the experimental evaluation. Section~\ref{sec:generalization} discusses generalization to other RAN functions. Related work is reviewed in Section~\ref{sec:related}. Finally, Section~\ref{sec:conclusion} concludes the paper.

\begin{figure*}[htbp]
  \centering
  \includegraphics[scale=0.48]{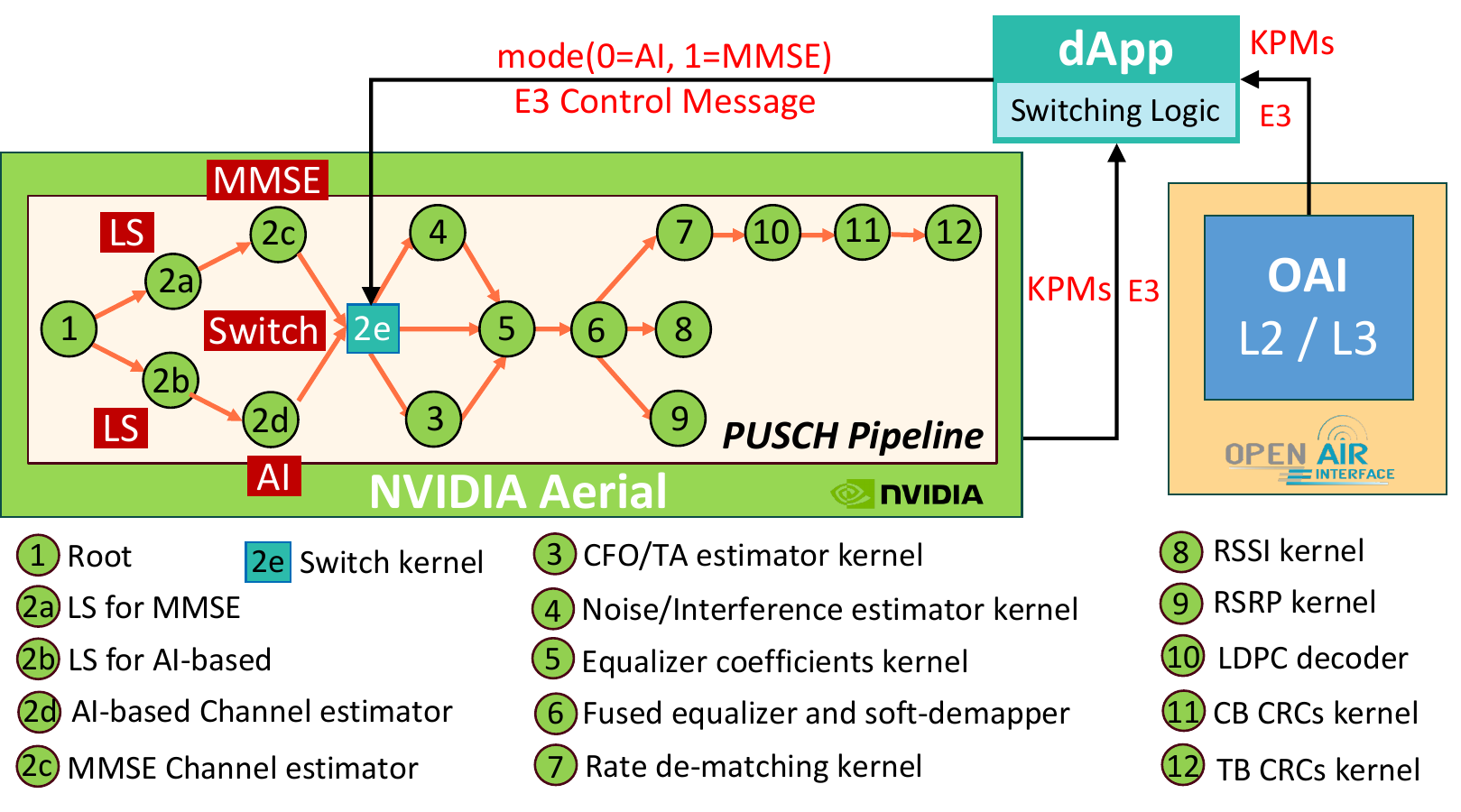}
    \setlength\belowcaptionskip{-.3cm} 
        \setlength\abovecaptionskip{0cm}
\caption{ARCHES prototype instantiated for channel estimation, 
illustrating NVIDIA Aerial, OAI, and the dApp. The switch kernel 
(node~2e) selects between Expert~A (MMSE, node~2c) and Expert~B 
(AI-based, node~2d). The dApp collects KPMs and communicates the 
\texttt{mode} variable via the E3 interface (Section~\ref{sec:dapp_control}). 
The same kernel and control path generalize to other PHY functions.}
  \label{fig:prototype}
\end{figure*}

\vspace{-.3cm}

\section{Design Principles for Real-Time Expert Switching in the RAN}
\label{sec:principles}

Before detailing the \switch framework, we state the design invariants that any real-time expert-switching mechanism for GPU-accelerated \gls{phy} pipelines must satisfy. These principles are independent of the specific signal processing function being switched and emerge from the constraints of inline GPU processing at sub-millisecond timescales.

\textbf{Dual execution modes balancing latency and efficiency.} \switch supports two execution modes. In concurrent mode, all candidate experts execute in parallel on every slot, ensuring the selected output is always readily available and eliminating cold-start penalties; this mode also enables online benchmarking by exposing all expert outputs simultaneously. In selected-only mode, only the active expert executes per slot, reducing compute and energy consumption at the cost of at least a one-slot activation delay when switching. The choice between modes reflects a deployment-specific trade-off: concurrent mode prioritizes switching responsiveness and observability, while selected-only mode prioritizes resource efficiency.

\textbf{Uniform downstream interface via memory aliasing.} Downstream processing stages (equalization, demapping, decoding) read from a single, fixed memory buffer. The switch kernel either performs a no-op or copies the alternative expert's output into that buffer, depending on which expert is active. Downstream modules therefore require no modification regardless of how many experts are added.

\textbf{Slot-boundary timing semantics.} Switching decisions take effect at slot boundaries. A decision generated during slot $n$ is applied at the setup phase of slot $n+1$. Because timing is deterministic, no in-flight data is corrupted mid-slot and the control loop has a well-defined, predictable latency.

\textbf{Fail-safe defaults.} The switch defaults to the conventional (non-\gls{ai}) expert at initialization and on dApp failure. The system therefore never depends on the availability of either the \gls{ai} model or the dApp control plane for baseline operation.

\textbf{External control plane with independent policy evolution.} The switching policy resides in a dApp external to the real-time pipeline. Because the policy lives outside the pipeline, it can be updated, retrained, or replaced without modifying the \gls{phy} pipeline or its CUDA graphs. The dApp communicates only a single scalar (the \texttt{mode} variable) to the pipeline per slot.

\textbf{Cross-layer telemetry as policy input.} The switching decision is informed by \glspl{kpm} collected across multiple protocol layers, from \gls{phy}-layer channel quality metrics to \gls{mac}-layer scheduling and throughput indicators. No single metric suffices to capture the multi-dimensional trade-off between estimation quality, compute cost, and throughput gain.

\vspace{-.4cm}

\section{The \switch Framework}
\label{sec:framework}

This section describes the \switch framework, 
independent of the specific signal processing function being 
switched. \switch has three main components, illustrated 
in Figure~\ref{fig:sys_archi}: (i)~a switchable expert bank 
integrated into the \gls{phy} pipeline, where $N$ experts can
execute on the same input and a CUDA switch kernel 
selects the designated output; (ii)~an external switching logic 
that determines which expert best suits the operating regime; and (iii)~a dApp-based control path that collects 
cross-layer telemetry and propagates the switching decision to 
the pipeline.

The system architecture builds on the O-RAN disaggregated 
\gls{gnb} model, which separates base station functionality into 
modular, programmable components connected via open interfaces: 
the \gls{ru}, \gls{du}, and \gls{cu}. \switch is built on this 
architecture, where NVIDIA Aerial~\cite{arc-ota} implements high-\gls{phy} 
processing and \gls{oai} manages Layer~2 and higher layers. We use NVIDIA Aerial \gls{cubb}, a \gls{sdk} that 
handles Low-DU/High-\gls{phy} signal processing for \glspl{gpu}, 
implemented in CUDA and C++. \gls{cubb} operates with slot-level 
granularity, where each slot has a duration of 500~$\mu$s 
(30~kHz subcarrier spacing). Inline \gls{gpu} acceleration is what makes \switch possible: multiple experts run within the same pipeline, and the switching decision is applied at the output selection stage without interrupting the processing flow.

Figure~\ref{fig:prototype} shows the detailed prototype 
instantiated for channel estimation: the \gls{pusch} CUDA graph 
executes both experts in parallel (nodes~2c and~2d), with the 
switch kernel (node~2e) selecting the output, while the dApp 
collects \glspl{kpm} from both Aerial and \gls{oai} and 
communicates the \texttt{mode} variable back to Aerial.

\vspace{-.3cm}
\subsection{Switchable Expert Bank}
\label{sec:expert_bank}

The switchable expert bank resides within the Low-DU (High-\gls{phy}), where it replaces a single processing module with a bank of $N$ experts ($\text{Expert}_1, \text{Expert}_2, \ldots, \text{Expert}_N$) that execute on the same input data. \switch supports concurrent execution of all experts for two reasons: (i)~it eliminates switching latency, since the selected output is always readily available regardless of when the switch logic updates its decision, and (ii)~it enables the pipeline to expose intermediate outputs from all experts, which helps monitor online performance and benchmark the switching policy offline. Each expert produces its own output, and a CUDA switch kernel selects the designated result, forwarding it to downstream processing stages.

Alternatively, \switch supports a selected-only execution mode in which only the active expert runs per slot, reducing compute and energy consumption at the cost of at least a one-slot activation delay when switching. In this work, we use concurrent execution to validate seamless switching and quantify expert-level tradeoffs.

\vspace{-.3cm}
\subsection{CUDA Switch Kernel}
\label{sec:switch_kernel}

The switch kernel implements the runtime selection logic on the 
\gls{gpu}. Its behavior is controlled by a binary variable, 
\texttt{mode}, assigned during the slot setup phase. By 
default, \texttt{mode} is initialized to 1 (\gls{mmse}), ensuring 
fail-safe operation before the dApp issues its first decision or 
in case of dApp failure.

To support efficient switching, separate memory locations are 
allocated for each expert's output. Downstream modules 
consistently read from the AI-designated memory buffer. When 
\texttt{mode}=0 (AI), the kernel performs no operation, as the 
AI output is already in place. When \texttt{mode}=1, the 
kernel copies the \gls{mmse} output into the AI buffer using coalesced 
memory access. This memory layout choice means the \gls{mmse} path 
incurs a small copy overhead, while the AI path incurs none, a deliberate trade-off, since the AI-based estimator already dominates the latency budget 
and the additional copy overhead is better absorbed by the 
lighter \gls{mmse} path. We profile the execution times in Section~\ref{sec:latency}.

\vspace{-.3cm}
\subsection{dApp-Based Control Path}
\label{sec:dapp_control}

The switching logic resides in a dApp, a lightweight application deployed directly on the \gls{gnb} that can access \gls{phy}/\gls{mac} telemetry and execute control actions at sub-10~ms timescales~\cite{D_Oro_2022}. The \switch dApp collects relevant \glspl{kpm} from both 
NVIDIA Aerial and \gls{oai} via the E3 interface, processes 
them through a switching policy, and determines the optimal expert. The selected expert is encoded in the \texttt{mode} variable and conveyed to the \gls{phy} pipeline to trigger the corresponding switch.

We leverage the dApp framework~\cite{lacava2025dapps} implemented in the NVIDIA Aerial stack~\cite{villa2025cusense}. dApps are real-time, co-located applications interfacing directly with the \gls{gnb} through a structured pre-standard E3-based control and data plane. This provides low-latency access to fine-grained \gls{phy}/\gls{mac} telemetry and optional control over RAN behavior. At the interface level, the E3 Agent within the RAN exposes telemetry (e.g., channel estimates, I/Q samples, and metadata) via shared memory and lightweight indication messages, while the E3 Manager within the dApp handles setup, subscription, and data delivery to the application logic.

Internally, the dApp has three main components: (i) an orchestration layer (E3 Manager), (ii) a processing layer implementing the dApp logic, and (iii) a client interface for lifecycle management. Multiple inference engines and AI backends can therefore be integrated without structural changes. At runtime, data flows from the CUDA-based L1 pipeline to pinned host memory and shared-memory buffers, giving co-located dApps near zero-copy access while preserving isolation and real-time performance.

\textbf{dApp--Switch Kernel Interaction.} The control path from the dApp to the \gls{gpu} follows a structured sequence:
\emph{Decision and signaling:} The dApp encodes its expert selection into the \texttt{mode} variable and transmits it to NVIDIA Aerial via the E3 interface. This value is made available on the host (CPU) side.
\emph{Control integration:} The \texttt{mode} parameter is incorporated into the derived \gls{ue} group parameters during the setup phase of each \gls{ul} slot.
\emph{Host-to-device propagation:} The \texttt{mode} parameter is copied from host memory to \gls{gpu} memory together with the rest of the processing context.
\emph{Kernel-level execution:} On the \gls{gpu}, the switch kernel reads \texttt{mode} and selects the corresponding expert output.
\emph{Timing semantics:} Mode updates take effect at slot boundaries. A value generated during slot $n$ is applied at slot $n+1$; mid-slot updates are deferred.


\vspace{-.3cm}
\section{Methodology for Policy Design}
\label{sec:methodology}

Given a bank of experts, the next question is how to select telemetry inputs and design the switching policy. We design the switching policy in three stages that are transferable across different \gls{phy} functions. We demonstrate this through the channel estimation case study, but the stages themselves --- controlled perturbation, monotonicity filtering, and redundancy reduction --- are generic.

\begin{figure}[htbp]
  \centering
      \includegraphics[width=1\columnwidth]{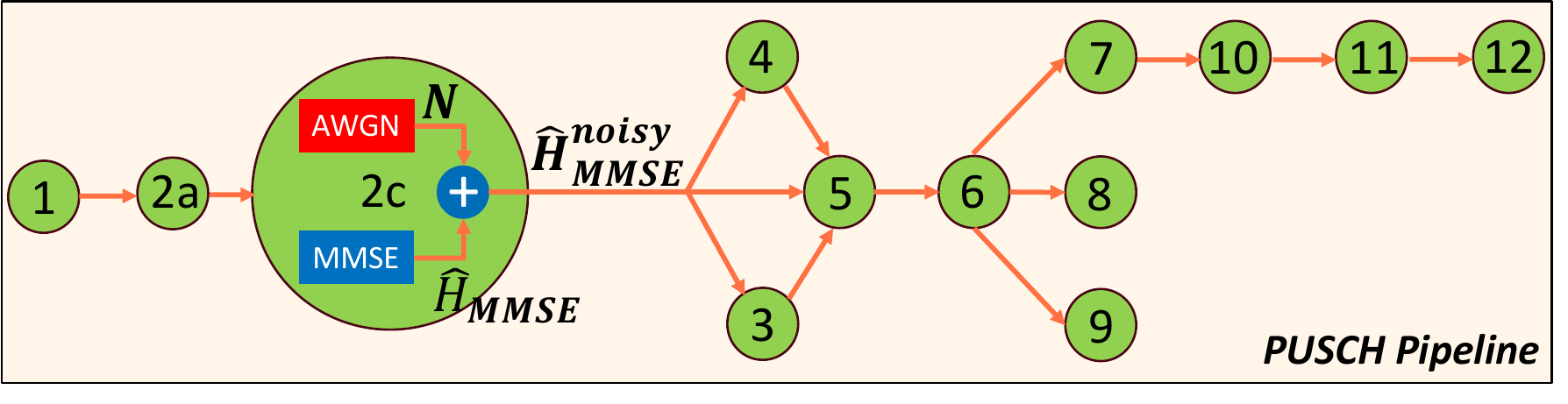}
  \setlength\abovecaptionskip{-.3cm}
  \caption{Controlled perturbation for KPM sensitivity analysis: noise injection into the channel estimates within Expert~A (node~2c), reusing the node descriptions from Figure~\ref{fig:prototype}.}
  \label{fig:noise}
\end{figure}

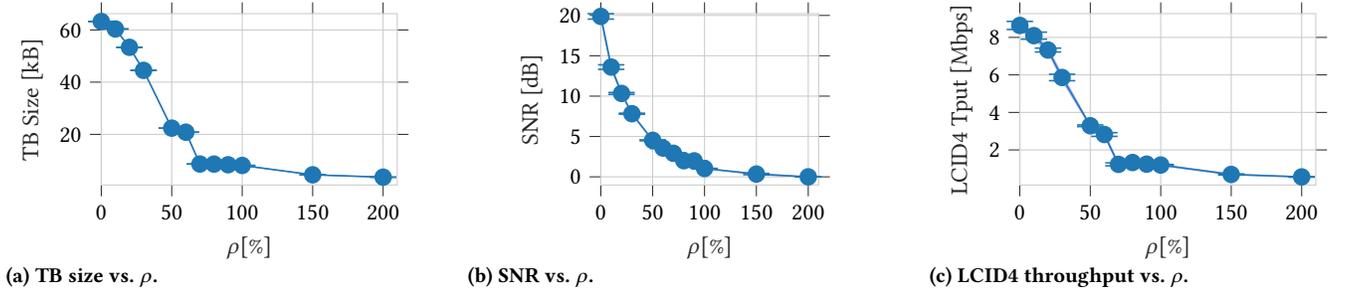
\begin{figure*}[htbp]
    \centering
    \begin{subfigure}[b]{0.31\textwidth}
        \centering
        \setlength\fheight{0.35\linewidth}
\begin{tikzpicture}

\definecolor{black38}{RGB}{38,38,38}
\definecolor{gray31119180}{RGB}{31,119,180}
\definecolor{white204}{RGB}{204,204,204}

\begin{axis}[
    height=.7\textwidth,         
    width=\textwidth,   
axis line style={white204},
tick align=outside,
x grid style={white204},
xlabel=\textcolor{black38}{$\rho$[\%]},
xmajorgrids,
xmajorticks=true,
xmin=0, xmax=210,
xtick style={color=black38},
y grid style={white204},
ylabel=\textcolor{black38}{TB Size [kB]},
ymajorgrids,
ymajorticks=true,
ymin=0.595039512540177, ymax=66.2441033829402,
ytick style={color=black38}
]
\path [draw=blue, fill=blue, opacity=0.3]
(axis cs:0,63.2600550251947)
--(axis cs:0,63.2296253952178)
--(axis cs:10,60.3573849732019)
--(axis cs:20,53.3345738493898)
--(axis cs:30,44.494923281864)
--(axis cs:50,22.3567478108459)
--(axis cs:60,20.8082715797603)
--(axis cs:70,8.58817161487772)
--(axis cs:80,8.57494947250893)
--(axis cs:90,8.3266085252222)
--(axis cs:100,8.06017660218782)
--(axis cs:150,4.481596273821)
--(axis cs:200,3.57908787028563)
--(axis cs:200,3.60494044176238)
--(axis cs:200,3.60494044176238)
--(axis cs:150,4.51629508839718)
--(axis cs:100,8.10331897761384)
--(axis cs:90,8.37090995310543)
--(axis cs:80,8.61755020968099)
--(axis cs:70,8.66480527209005)
--(axis cs:60,20.8481675697103)
--(axis cs:50,22.3957181522288)
--(axis cs:30,44.5319709219223)
--(axis cs:20,53.3562168104589)
--(axis cs:10,60.3893524354702)
--(axis cs:0,63.2600550251947)
--cycle;

\path [draw=gray31119180, thick]
(axis cs:0,63.2296253952178)
--(axis cs:0,63.2600550251947);

\path [draw=gray31119180, thick]
(axis cs:10,60.3573849732019)
--(axis cs:10,60.3893524354702);

\path [draw=gray31119180, thick]
(axis cs:20,53.3345738493898)
--(axis cs:20,53.3562168104589);

\path [draw=gray31119180, thick]
(axis cs:30,44.494923281864)
--(axis cs:30,44.5319709219223);

\path [draw=gray31119180, thick]
(axis cs:50,22.3567478108459)
--(axis cs:50,22.3957181522288);

\path [draw=gray31119180, thick]
(axis cs:60,20.8082715797603)
--(axis cs:60,20.8481675697103);

\path [draw=gray31119180, thick]
(axis cs:70,8.58817161487772)
--(axis cs:70,8.66480527209005);

\path [draw=gray31119180, thick]
(axis cs:80,8.57494947250893)
--(axis cs:80,8.61755020968099);

\path [draw=gray31119180, thick]
(axis cs:90,8.3266085252222)
--(axis cs:90,8.37090995310543);

\path [draw=gray31119180, thick]
(axis cs:100,8.06017660218782)
--(axis cs:100,8.10331897761384);

\path [draw=gray31119180, thick]
(axis cs:150,4.481596273821)
--(axis cs:150,4.51629508839718);

\path [draw=gray31119180, thick]
(axis cs:200,3.57908787028563)
--(axis cs:200,3.60494044176238);

\addplot [semithick, gray31119180, mark=-, mark size=5, mark options={solid}, only marks]
table {%
0 63.2296253952178
10 60.3573849732019
20 53.3345738493898
30 44.494923281864
50 22.3567478108459
60 20.8082715797603
70 8.58817161487772
80 8.57494947250893
90 8.3266085252222
100 8.06017660218782
150 4.481596273821
200 3.57908787028563
};
\addplot [semithick, gray31119180, mark=-, mark size=5, mark options={solid}, only marks]
table {%
0 63.2600550251947
10 60.3893524354702
20 53.3562168104589
30 44.5319709219223
50 22.3957181522288
60 20.8481675697103
70 8.66480527209005
80 8.61755020968099
90 8.37090995310543
100 8.10331897761384
150 4.51629508839718
200 3.60494044176238
};
\addplot [semithick, gray31119180, mark=*, mark size=3, mark options={solid}]
table {%
0 63.2448402102063
10 60.3733687043361
20 53.3453953299243
30 44.5134471018931
50 22.3762329815373
60 20.8282195747353
70 8.62648844348389
80 8.59624984109496
90 8.34875923916382
100 8.08174778990083
150 4.49894568110909
200 3.592014156024
};
\end{axis}

\end{tikzpicture}
         \captionsetup{aboveskip=-1pt}
        \caption{TB size vs.\ $\rho$.}
        \label{fig:KPM_tbsize}
    \end{subfigure}
    \hfill
    \begin{subfigure}[b]{0.31\textwidth}
        \centering
        \setlength\fheight{0.35\linewidth}
\begin{tikzpicture}

\definecolor{black38}{RGB}{38,38,38}
\definecolor{gray31119180}{RGB}{31,119,180}
\definecolor{white204}{RGB}{204,204,204}

\begin{axis}[
    height=.7\textwidth,         
axis line style={white204},
tick align=outside,
x grid style={white204},
xlabel=\textcolor{black38}{$\rho$[\%]},
xmajorgrids,
xmajorticks=true,
xmin=0, xmax=210,
xtick style={color=black38},
y grid style={white204},
ylabel=\textcolor{black38}{SNR [dB]},
ymajorgrids,
ymajorticks=true,
ymin=-1.01057376924304, ymax=20.2220491541039,
ytick style={color=black38}
]
\path [draw=blue, fill=blue, opacity=0.3]
(axis cs:0,20.2114753848609)
--(axis cs:0,19.5461003727149)
--(axis cs:10,13.3189337455556)
--(axis cs:20,10.2196895276295)
--(axis cs:30,7.75896440900268)
--(axis cs:50,4.4658929886753)
--(axis cs:60,3.49918095449646)
--(axis cs:70,2.86493539476426)
--(axis cs:80,1.99098108563329)
--(axis cs:90,1.96150561949989)
--(axis cs:100,1.01135633290223)
--(axis cs:150,0.265495418690024)
--(axis cs:200,0)
--(axis cs:200,0)
--(axis cs:200,0)
--(axis cs:150,0.42324630316428)
--(axis cs:100,1.05292938138349)
--(axis cs:90,1.99563723764297)
--(axis cs:80,2.00293988701108)
--(axis cs:70,2.96992699055685)
--(axis cs:60,3.62485005325548)
--(axis cs:50,4.58473992271711)
--(axis cs:30,7.88589540408143)
--(axis cs:20,10.4469771390372)
--(axis cs:10,13.8840182839646)
--(axis cs:0,20.2114753848609)
--cycle;

\path [draw=gray31119180, thick]
(axis cs:0,19.5461003727149)
--(axis cs:0,20.2114753848609);

\path [draw=gray31119180, thick]
(axis cs:10,13.3189337455556)
--(axis cs:10,13.8840182839646);

\path [draw=gray31119180, thick]
(axis cs:20,10.2196895276295)
--(axis cs:20,10.4469771390372);

\path [draw=gray31119180, thick]
(axis cs:30,7.75896440900268)
--(axis cs:30,7.88589540408143);

\path [draw=gray31119180, thick]
(axis cs:50,4.4658929886753)
--(axis cs:50,4.58473992271711);

\path [draw=gray31119180, thick]
(axis cs:60,3.49918095449646)
--(axis cs:60,3.62485005325548);

\path [draw=gray31119180, thick]
(axis cs:70,2.86493539476426)
--(axis cs:70,2.96992699055685);

\path [draw=gray31119180, thick]
(axis cs:80,1.99098108563329)
--(axis cs:80,2.00293988701108);

\path [draw=gray31119180, thick]
(axis cs:90,1.96150561949989)
--(axis cs:90,1.99563723764297);

\path [draw=gray31119180, thick]
(axis cs:100,1.01135633290223)
--(axis cs:100,1.05292938138349);

\path [draw=gray31119180, thick]
(axis cs:150,0.265495418690024)
--(axis cs:150,0.42324630316428);

\path [draw=gray31119180, thick]
(axis cs:200,0)
--(axis cs:200,0);

\addplot [semithick, gray31119180, mark=-, mark size=5, mark options={solid}, only marks]
table {%
0 19.5461003727149
10 13.3189337455556
20 10.2196895276295
30 7.75896440900268
50 4.4658929886753
60 3.49918095449646
70 2.86493539476426
80 1.99098108563329
90 1.96150561949989
100 1.01135633290223
150 0.265495418690024
200 0
};
\addplot [semithick, gray31119180, mark=-, mark size=5, mark options={solid}, only marks]
table {%
0 20.2114753848609
10 13.8840182839646
20 10.4469771390372
30 7.88589540408143
50 4.58473992271711
60 3.62485005325548
70 2.96992699055685
80 2.00293988701108
90 1.99563723764297
100 1.05292938138349
150 0.42324630316428
200 0
};
\addplot [semithick, gray31119180, mark=*, mark size=3, mark options={solid}]
table {%
0 19.8787878787879
10 13.6014760147601
20 10.3333333333333
30 7.82242990654206
50 4.5253164556962
60 3.56201550387597
70 2.91743119266055
80 1.99696048632219
90 1.97857142857143
100 1.03214285714286
150 0.344370860927152
200 0
};
\end{axis}

\end{tikzpicture}
         \captionsetup{aboveskip=-1pt}
        \caption{SNR vs.\ $\rho$.}
        \label{fig:KPM_snr}
    \end{subfigure}
    \hfill
    \begin{subfigure}[b]{0.31\textwidth}
        \centering
        \setlength\fheight{0.35\linewidth}
\begin{tikzpicture}

\definecolor{black38}{RGB}{38,38,38}
\definecolor{gray31119180}{RGB}{31,119,180}
\definecolor{white204}{RGB}{204,204,204}

\begin{axis}[
    height=.7\textwidth,         
    width=\textwidth,   
axis line style={white204},
tick align=outside,
x grid style={white204},
xlabel=\textcolor{black38}{$\rho$[\%]},
xmajorgrids,
xmajorticks=true,
xmin=0, xmax=210,
xtick style={color=black38},
y grid style={white204},
ylabel=\textcolor{black38}{LCID4 Tput [Mbps]},
ymajorgrids,
ymajorticks=true,
ymin=0.12992778931932, ymax=9.26688245620514,
ytick style={color=black38}
]
\path [draw=blue, fill=blue, opacity=0.3]
(axis cs:0,8.85156633498306)
--(axis cs:0,8.42157185551573)
--(axis cs:10,7.90616645512625)
--(axis cs:20,7.22206649431242)
--(axis cs:30,5.68979435668801)
--(axis cs:50,3.24330154937923)
--(axis cs:60,2.72177647144254)
--(axis cs:70,1.15217049985244)
--(axis cs:80,1.30517411872829)
--(axis cs:90,1.21340792430892)
--(axis cs:100,1.16299845889533)
--(axis cs:150,0.664298366516866)
--(axis cs:200,0.545243910541409)
--(axis cs:200,0.575952059329649)
--(axis cs:200,0.575952059329649)
--(axis cs:150,0.72056690104792)
--(axis cs:100,1.22417322465626)
--(axis cs:90,1.270684770941)
--(axis cs:80,1.35135217242403)
--(axis cs:70,1.3078890098037)
--(axis cs:60,2.92068573435153)
--(axis cs:50,3.33323269063953)
--(axis cs:30,6.03408595220647)
--(axis cs:20,7.4278909007855)
--(axis cs:10,8.27921080056522)
--(axis cs:0,8.85156633498306)
--cycle;

\path [draw=gray31119180, thick]
(axis cs:0,8.42157185551573)
--(axis cs:0,8.85156633498306);

\path [draw=gray31119180, thick]
(axis cs:10,7.90616645512625)
--(axis cs:10,8.27921080056522);

\path [draw=gray31119180, thick]
(axis cs:20,7.22206649431242)
--(axis cs:20,7.4278909007855);

\path [draw=gray31119180, thick]
(axis cs:30,5.68979435668801)
--(axis cs:30,6.03408595220647);

\path [draw=gray31119180, thick]
(axis cs:50,3.24330154937923)
--(axis cs:50,3.33323269063953);

\path [draw=gray31119180, thick]
(axis cs:60,2.72177647144254)
--(axis cs:60,2.92068573435153);

\path [draw=gray31119180, thick]
(axis cs:70,1.15217049985244)
--(axis cs:70,1.3078890098037);

\path [draw=gray31119180, thick]
(axis cs:80,1.30517411872829)
--(axis cs:80,1.35135217242403);

\path [draw=gray31119180, thick]
(axis cs:90,1.21340792430892)
--(axis cs:90,1.270684770941);

\path [draw=gray31119180, thick]
(axis cs:100,1.16299845889533)
--(axis cs:100,1.22417322465626);

\path [draw=gray31119180, thick]
(axis cs:150,0.664298366516866)
--(axis cs:150,0.72056690104792);

\path [draw=gray31119180, thick]
(axis cs:200,0.545243910541409)
--(axis cs:200,0.575952059329649);

\addplot [semithick, gray31119180, mark=-, mark size=5, mark options={solid}, only marks]
table {%
0 8.42157185551573
10 7.90616645512625
20 7.22206649431242
30 5.68979435668801
50 3.24330154937923
60 2.72177647144254
70 1.15217049985244
80 1.30517411872829
90 1.21340792430892
100 1.16299845889533
150 0.664298366516866
200 0.545243910541409
};
\addplot [semithick, gray31119180, mark=-, mark size=5, mark options={solid}, only marks]
table {%
0 8.85156633498306
10 8.27921080056522
20 7.4278909007855
30 6.03408595220647
50 3.33323269063953
60 2.92068573435153
70 1.3078890098037
80 1.35135217242403
90 1.270684770941
100 1.22417322465626
150 0.72056690104792
200 0.575952059329649
};
\addplot [semithick, gray31119180, mark=*, mark size=3, mark options={solid}]
table {%
0 8.63656909524939
10 8.09268862784574
20 7.32497869754896
30 5.86194015444724
50 3.28826712000938
60 2.82123110289704
70 1.23002975482807
80 1.32826314557616
90 1.24204634762496
100 1.1935858417758
150 0.692432633782393
200 0.560597984935529
};
\end{axis}

\end{tikzpicture}
         \captionsetup{aboveskip=-1pt}
        \caption{LCID4 throughput vs.\ $\rho$.}
        \label{fig:KPM_lcid4}
    \end{subfigure}
    \captionsetup{aboveskip=0pt, belowskip=-8pt}
    \caption{KPM degradation trends showing incremental performance loss with increasing perturbation intensity $\rho$ for the channel estimation case study.}
    \label{fig:KPM_trends}
\end{figure*}

\vspace{-.4cm}
\subsection{Stage 1: Sensitivity Analysis via Controlled Perturbation}
\label{sec:stage1}

The first stage identifies candidate \glspl{kpm} that are 
sensitive to the quality of the processing function under study. 
The approach is to inject controlled, calibrated degradation into 
one expert's output and observe which downstream metrics respond. 
By isolating the effect of processing quality from the switching 
logic itself, we ensure that the selected \glspl{kpm} genuinely 
reflect processing accuracy rather than artifacts of the 
switching process. Note that the \switch switching mechanism and 
the AI-based estimator are not included, as this setup is 
designed solely to serve as a baseline platform for KPM 
extraction.

For the channel estimation case study, we inject \gls{awgn} into 
the \gls{mmse} channel estimates at node~2c, as illustrated in Figure~\ref{fig:noise}. The wireless channel is characterized by the \gls{csi} 
$\mathbf{H} \in \mathbb{C}^{N_{\mathrm{ant}} \times N_{\mathrm{l}} \times 
N_{\mathrm{sc}} \times N_{\mathrm{sym}}}$, where $N_{\mathrm{ant}} = 4$ is the 
number of antenna ports, $N_{\mathrm{l}} = 1$ is the number of transmission 
layers, $N_{\mathrm{sc}} = 12 \times N_{\mathrm{PRB}}$ is the number of 
subcarriers with $N_{\mathrm{PRB}}$ denoting the number of \glspl{prb} 
allocated for \gls{ul} transmission, and $N_{\mathrm{sym}}=14$ is the number 
of \gls{ofdm} symbols in a slot. At the receiver, $\mathbf{H}$ is estimated 
using \gls{dmrs}-based \gls{mmse} estimation with frequency-domain interpolation at 
\gls{dmrs} positions, yielding
$\hat{\mathbf{H}}_{\mathrm{MMSE}} \in \mathbb{C}^{N_{\mathrm{ant}} \times 
N_{\mathrm{l}} \times N_{\mathrm{sc}} \times 
N_{\mathrm{sym}}^{\mathrm{DMRS}}}$,
where estimates are obtained from $N_{\mathrm{sym}}^{\mathrm{DMRS}} = 3$ 
pilot symbols and later interpolated across all \gls{ofdm} symbols by the 
equalizer.

To induce degraded channel knowledge for \gls{ota} transmissions, controlled 
synthetic noise $\mathbf{N}$ of identical dimensions is injected into 
$\hat{\mathbf{H}}_{\mathrm{MMSE}}$, modeled as complex \gls{awgn}:
\begin{equation}
    \mathbf{N} \sim \mathcal{CN}(0, \sigma^2),
\end{equation}
where the standard deviation is proportional to the mean magnitude of 
$\hat{\mathbf{H}}_{\mathrm{MMSE}}$:
\begin{equation}
    \sigma = \rho \cdot \mathbb{E}\!\left[ |\hat{\mathbf{H}}_{\mathrm{MMSE}}| 
    \right].
\end{equation}

The parameter $\rho \in [0, 2]$ controls the noise intensity. We sweep 
$\rho$ from 0 to 2 in steps of 0.1 to evaluate performance degradation under 
gradually increasing noise levels. The resulting corrupted channel estimate 
$\hat{\mathbf{H}}_{\mathrm{MMSE}}^{\mathrm{noisy}}$ is therefore given by


\begin{equation}
    \hat{\mathbf{H}}_{\mathrm{MMSE}}^{\mathrm{noisy}} 
        = \hat{\mathbf{H}}_{\mathrm{MMSE}} + \rho \cdot \mathbb{E}\!\left[
           |\hat{\mathbf{H}}_{\mathrm{MMSE}}| \right] \cdot 
           \mathcal{CN}(0,1),
\end{equation}

as illustrated in Figure~\ref{fig:noise}.

We then evaluate the degradation of candidate \glspl{kpm} derived from 
NVIDIA Aerial Data Lake, described in Section~\ref{sec:eval} and \gls{oai} as a function of $\rho$, identifying those that exhibit 
consistent sensitivity to channel estimation errors.

For other functions, the perturbation would differ: for beam management, one might inject perturbations into channel estimates or angular offsets into beam predictions; for \gls{ldpc} decoding variants, one might degrade decoder inputs by injecting noise into log-likelihood ratios at controlled rates. The idea is the same: perturb the expert output and observe how downstream performance changes.


\vspace{-.3cm}

\subsection{Stage 2: Monotonicity Filtering}
\label{sec:stage2}

Having identified candidate \glspl{kpm} via perturbation, the second stage retains only those that exhibit a consistent and monotonic degradation trend as perturbation intensity ($\rho$) increases, indicating a direct sensitivity to processing quality.

As shown in Figure~\ref{fig:KPM_trends}, all three representative \glspl{kpm} exhibit a clear 
monotonic decline as $\rho$ increases. \gls{tb} size 
(Figure~\ref{fig:KPM_tbsize}) and \gls{snr} 
(Figure~\ref{fig:KPM_snr}) follow a similar trajectory: both remain 
relatively stable at low noise levels before dropping sharply in the 
mid-range of $\rho$, reflecting the point at which channel estimation 
errors begin to significantly impact link adaptation and demodulation. 
LCID4 throughput (Figure~\ref{fig:KPM_lcid4}) follows an analogous 
trend at the \gls{mac} layer, confirming that the \gls{phy}-layer degradation 
propagates upward through the protocol stack. The narrow 95\% confidence 
intervals across all three plots indicate that these trends are 
consistent and repeatable across experimental runs.

\vspace{-.3cm}
\subsection{Stage 3: Redundancy Reduction via Hierarchical Clustering}
\label{sec:stage3}


\begin{figure}[htbp]
    \centering
\begin{subfigure}[b]{.6\columnwidth}
    \centering
    \resizebox{\linewidth}{!}{\definecolor{c10}{RGB}{5,48,97}
\definecolor{c09}{RGB}{33,102,172}
\definecolor{c08}{RGB}{67,147,195}
\definecolor{c07}{RGB}{146,197,222}
\definecolor{c06}{RGB}{209,229,240}
\definecolor{c05}{RGB}{247,247,247}
\definecolor{c04}{RGB}{253,219,199}
\definecolor{c03}{RGB}{244,165,130}
\definecolor{c02}{RGB}{214,96,77}
\definecolor{c01}{RGB}{178,24,43}
\definecolor{c00}{RGB}{103,0,31}
\begin{tikzpicture}
    \fill[fill=c10] (-.5,7.5) rectangle (.5,8.5);  \node[font=\huge, text=white] at (0,8) {1};
    \fill[fill=c08] (.5,7.5) rectangle (1.5,8.5);   \node[font=\huge, text=white] at (1,8) {.74};
    \fill[fill=c08] (1.5,7.5) rectangle (2.5,8.5);   \node[font=\huge, text=white] at (2,8) {.75};
    \fill[fill=c08] (2.5,7.5) rectangle (3.5,8.5);   \node[font=\huge, text=white] at (3,8) {.76};
    \fill[fill=c08] (3.5,7.5) rectangle (4.5,8.5);   \node[font=\huge, text=white] at (4,8) {.79};
    \fill[fill=c09] (4.5,7.5) rectangle (5.5,8.5);   \node[font=\huge, text=white] at (5,8) {.8};
    \fill[fill=c09] (5.5,7.5) rectangle (6.5,8.5);   \node[font=\huge, text=white] at (6,8) {.8};
    \fill[fill=c04] (6.5,7.5) rectangle (7.5,8.5);   \node[font=\huge] at (7,8) {-.32};
    \fill[fill=c02] (7.5,7.5) rectangle (8.5,8.5);   \node[font=\huge] at (8,8) {-.58};

    \fill[fill=c08] (-.5,6.5) rectangle (.5,7.5);  \node[font=\huge, text=white] at (0,7) {.74};
    \fill[fill=c10] (.5,6.5) rectangle (1.5,7.5);   \node[font=\huge, text=white] at (1,7) {1};
    \fill[fill=c09] (1.5,6.5) rectangle (2.5,7.5);   \node[font=\huge, text=white] at (2,7) {.83};
    \fill[fill=c09] (2.5,6.5) rectangle (3.5,7.5);   \node[font=\huge, text=white] at (3,7) {.81};
    \fill[fill=c09] (3.5,6.5) rectangle (4.5,7.5);   \node[font=\huge, text=white] at (4,7) {.87};
    \fill[fill=c09] (4.5,6.5) rectangle (5.5,7.5);   \node[font=\huge, text=white] at (5,7) {.91};
    \fill[fill=c09] (5.5,6.5) rectangle (6.5,7.5);   \node[font=\huge, text=white] at (6,7) {.91};
    \fill[fill=c04] (6.5,6.5) rectangle (7.5,7.5);   \node[font=\huge] at (7,7) {-.23};
    \fill[fill=c01] (7.5,6.5) rectangle (8.5,7.5);   \node[font=\huge, text=white] at (8,7) {-.72};

    \fill[fill=c08] (-.5,5.5) rectangle (.5,6.5);  \node[font=\huge, text=white] at (0,6) {.75};
    \fill[fill=c09] (.5,5.5) rectangle (1.5,6.5);   \node[font=\huge, text=white] at (1,6) {.83};
    \fill[fill=c10] (1.5,5.5) rectangle (2.5,6.5);   \node[font=\huge, text=white] at (2,6) {1};
    \fill[fill=c09] (2.5,5.5) rectangle (3.5,6.5);   \node[font=\huge, text=white] at (3,6) {.85};
    \fill[fill=c09] (3.5,5.5) rectangle (4.5,6.5);   \node[font=\huge, text=white] at (4,6) {.93};
    \fill[fill=c09] (4.5,5.5) rectangle (5.5,6.5);   \node[font=\huge, text=white] at (5,6) {.92};
    \fill[fill=c09] (5.5,5.5) rectangle (6.5,6.5);   \node[font=\huge, text=white] at (6,6) {.93};
    \fill[fill=c05] (6.5,5.5) rectangle (7.5,6.5);   \node[font=\huge] at (7,6) {-.19};
    \fill[fill=c02] (7.5,5.5) rectangle (8.5,6.5);   \node[font=\huge] at (8,6) {-.64};

    \fill[fill=c08] (-.5,4.5) rectangle (.5,5.5);  \node[font=\huge, text=white] at (0,5) {.76};
    \fill[fill=c09] (.5,4.5) rectangle (1.5,5.5);   \node[font=\huge, text=white] at (1,5) {.81};
    \fill[fill=c09] (1.5,4.5) rectangle (2.5,5.5);   \node[font=\huge, text=white] at (2,5) {.85};
    \fill[fill=c10] (2.5,4.5) rectangle (3.5,5.5);   \node[font=\huge, text=white] at (3,5) {1};
    \fill[fill=c09] (3.5,4.5) rectangle (4.5,5.5);   \node[font=\huge, text=white] at (4,5) {.95};
    \fill[fill=c10] (4.5,4.5) rectangle (5.5,5.5);   \node[font=\huge, text=white] at (5,5) {.97};
    \fill[fill=c10] (5.5,4.5) rectangle (6.5,5.5);   \node[font=\huge, text=white] at (6,5) {.96};
    \fill[fill=c04] (6.5,4.5) rectangle (7.5,5.5);   \node[font=\huge] at (7,5) {-.28};
    \fill[fill=c02] (7.5,4.5) rectangle (8.5,5.5);   \node[font=\huge] at (8,5) {-.69};

    \fill[fill=c08] (-.5,3.5) rectangle (.5,4.5);  \node[font=\huge, text=white] at (0,4) {.79};
    \fill[fill=c09] (.5,3.5) rectangle (1.5,4.5);   \node[font=\huge, text=white] at (1,4) {.87};
    \fill[fill=c09] (1.5,3.5) rectangle (2.5,4.5);   \node[font=\huge, text=white] at (2,4) {.93};
    \fill[fill=c09] (2.5,3.5) rectangle (3.5,4.5);   \node[font=\huge, text=white] at (3,4) {.95};
    \fill[fill=c10] (3.5,3.5) rectangle (4.5,4.5);   \node[font=\huge, text=white] at (4,4) {1};
    \fill[fill=c10] (4.5,3.5) rectangle (5.5,4.5);   \node[font=\huge, text=white] at (5,4) {.99};
    \fill[fill=c10] (5.5,3.5) rectangle (6.5,4.5);   \node[font=\huge, text=white] at (6,4) {.99};
    \fill[fill=c04] (6.5,3.5) rectangle (7.5,4.5);   \node[font=\huge] at (7,4) {-.25};
    \fill[fill=c02] (7.5,3.5) rectangle (8.5,4.5);   \node[font=\huge] at (8,4) {-.69};

    \fill[fill=c09] (-.5,2.5) rectangle (.5,3.5);  \node[font=\huge, text=white] at (0,3) {.8};
    \fill[fill=c09] (.5,2.5) rectangle (1.5,3.5);   \node[font=\huge, text=white] at (1,3) {.91};
    \fill[fill=c09] (1.5,2.5) rectangle (2.5,3.5);   \node[font=\huge, text=white] at (2,3) {.92};
    \fill[fill=c10] (2.5,2.5) rectangle (3.5,3.5);   \node[font=\huge, text=white] at (3,3) {.97};
    \fill[fill=c10] (3.5,2.5) rectangle (4.5,3.5);   \node[font=\huge, text=white] at (4,3) {.99};
    \fill[fill=c10] (4.5,2.5) rectangle (5.5,3.5);   \node[font=\huge, text=white] at (5,3) {1};
    \fill[fill=c10] (5.5,2.5) rectangle (6.5,3.5);   \node[font=\huge, text=white] at (6,3) {1};
    \fill[fill=c04] (6.5,2.5) rectangle (7.5,3.5);   \node[font=\huge] at (7,3) {-.26};
    \fill[fill=c01] (7.5,2.5) rectangle (8.5,3.5);   \node[font=\huge, text=white] at (8,3) {-.71};

    \fill[fill=c09] (-.5,1.5) rectangle (.5,2.5);  \node[font=\huge, text=white] at (0,2) {.8};
    \fill[fill=c09] (.5,1.5) rectangle (1.5,2.5);   \node[font=\huge, text=white] at (1,2) {.91};
    \fill[fill=c09] (1.5,1.5) rectangle (2.5,2.5);   \node[font=\huge, text=white] at (2,2) {.93};
    \fill[fill=c10] (2.5,1.5) rectangle (3.5,2.5);   \node[font=\huge, text=white] at (3,2) {.96};
    \fill[fill=c10] (3.5,1.5) rectangle (4.5,2.5);   \node[font=\huge, text=white] at (4,2) {.99};
    \fill[fill=c10] (4.5,1.5) rectangle (5.5,2.5);   \node[font=\huge, text=white] at (5,2) {1};
    \fill[fill=c10] (5.5,1.5) rectangle (6.5,2.5);   \node[font=\huge, text=white] at (6,2) {1};
    \fill[fill=c04] (6.5,1.5) rectangle (7.5,2.5);   \node[font=\huge] at (7,2) {-.26};
    \fill[fill=c01] (7.5,1.5) rectangle (8.5,2.5);   \node[font=\huge, text=white] at (8,2) {-.71};

    \fill[fill=c04] (-.5,.5) rectangle (.5,1.5);  \node[font=\huge] at (0,1) {-.32};
    \fill[fill=c04] (.5,.5) rectangle (1.5,1.5);   \node[font=\huge] at (1,1) {-.23};
    \fill[fill=c05] (1.5,.5) rectangle (2.5,1.5);   \node[font=\huge] at (2,1) {-.19};
    \fill[fill=c04] (2.5,.5) rectangle (3.5,1.5);   \node[font=\huge] at (3,1) {-.28};
    \fill[fill=c04] (3.5,.5) rectangle (4.5,1.5);   \node[font=\huge] at (4,1) {-.25};
    \fill[fill=c04] (4.5,.5) rectangle (5.5,1.5);   \node[font=\huge] at (5,1) {-.26};
    \fill[fill=c04] (5.5,.5) rectangle (6.5,1.5);   \node[font=\huge] at (6,1) {-.26};
    \fill[fill=c10] (6.5,.5) rectangle (7.5,1.5);   \node[font=\huge, text=white] at (7,1) {1};
    \fill[fill=c06] (7.5,.5) rectangle (8.5,1.5);   \node[font=\huge] at (8,1) {.16};

    \fill[fill=c02] (-.5,-.5) rectangle (.5,.5);  \node[font=\huge] at (0,0) {-.58};
    \fill[fill=c01] (.5,-.5) rectangle (1.5,.5);   \node[font=\huge, text=white] at (1,0) {-.72};
    \fill[fill=c02] (1.5,-.5) rectangle (2.5,.5);   \node[font=\huge] at (2,0) {-.64};
    \fill[fill=c02] (2.5,-.5) rectangle (3.5,.5);   \node[font=\huge] at (3,0) {-.69};
    \fill[fill=c02] (3.5,-.5) rectangle (4.5,.5);   \node[font=\huge] at (4,0) {-.69};
    \fill[fill=c01] (4.5,-.5) rectangle (5.5,.5);   \node[font=\huge, text=white] at (5,0) {-.71};
    \fill[fill=c01] (5.5,-.5) rectangle (6.5,.5);   \node[font=\huge, text=white] at (6,0) {-.71};
    \fill[fill=c06] (6.5,-.5) rectangle (7.5,.5);   \node[font=\huge] at (7,0) {.16};
    \fill[fill=c10] (7.5,-.5) rectangle (8.5,.5);   \node[font=\huge, text=white] at (8,0) {1};

    \node[font=\huge, rotate=45, anchor=east] at (0,-.7) {PDU Len};
    \node[font=\huge, rotate=45, anchor=east] at (1,-.7) {CodeRate};
    \node[font=\huge, rotate=45, anchor=east] at (2,-.7) {SINR};
    \node[font=\huge, rotate=45, anchor=east] at (3,-.7) {QAM Order};
    \node[font=\huge, rotate=45, anchor=east] at (4,-.7) {Num CB};
    \node[font=\huge, rotate=45, anchor=east] at (5,-.7) {MCS Index};
    \node[font=\huge, rotate=45, anchor=east] at (6,-.7) {TB Size};
    \node[font=\huge, rotate=45, anchor=east] at (7,-.7) {NDI};
    \node[font=\huge, rotate=45, anchor=east] at (8,-.7) {RSRP};

    \node[font=\huge, anchor=east] at (-.7,8) {PDU Len};
    \node[font=\huge, anchor=east] at (-.7,7) {CodeRate};
    \node[font=\huge, anchor=east] at (-.7,6) {SINR};
    \node[font=\huge, anchor=east] at (-.7,5) {QAM Order};
    \node[font=\huge, anchor=east] at (-.7,4) {Num CB};
    \node[font=\huge, anchor=east] at (-.7,3) {MCS Index};
    \node[font=\huge, anchor=east] at (-.7,2) {TB Size};
    \node[font=\huge, anchor=east] at (-.7,1) {NDI};
    \node[font=\huge, anchor=east] at (-.7,0) {RSRP};

    \draw (-.5,-.5) rectangle (8.5,8.5);
\end{tikzpicture}}
    \caption{DataLake cluster.}
    \label{fig:corr_datalake}
\end{subfigure}
\hfill
\begin{subfigure}[b]{.38\columnwidth}
    \centering
    \resizebox{\linewidth}{!}{\definecolor{c10}{RGB}{5,48,97}
\definecolor{c09}{RGB}{33,102,172}
\definecolor{c08}{RGB}{67,147,195}
\definecolor{c07}{RGB}{146,197,222}
\definecolor{c06}{RGB}{209,229,240}
\definecolor{c05}{RGB}{247,247,247}
\definecolor{c04}{RGB}{253,219,199}
\definecolor{c03}{RGB}{244,165,130}
\definecolor{c02}{RGB}{214,96,77}
\definecolor{c01}{RGB}{178,24,43}
\definecolor{c00}{RGB}{103,0,31}
\begin{tikzpicture}
    \fill[fill=c10] (-.5,3.5) rectangle (.5,4.5); \node[font=\huge, text=white] at (0,4) {1};
    \fill[fill=c06] (.5,3.5) rectangle (1.5,4.5);  \node[font=\huge] at (1,4) {.46};
    \fill[fill=c06] (1.5,3.5) rectangle (2.5,4.5);  \node[font=\huge] at (2,4) {.44};
    \fill[fill=c06] (2.5,3.5) rectangle (3.5,4.5);  \node[font=\huge] at (3,4) {.23};
    \fill[fill=c06] (3.5,3.5) rectangle (4.5,4.5);  \node[font=\huge] at (4,4) {.27};

    \fill[fill=c06] (-.5,2.5) rectangle (.5,3.5);  \node[font=\huge] at (0,3) {.46};
    \fill[fill=c10] (.5,2.5) rectangle (1.5,3.5);   \node[font=\huge, text=white] at (1,3) {1};
    \fill[fill=c08] (1.5,2.5) rectangle (2.5,3.5);   \node[font=\huge, text=white] at (2,3) {.7};
    \fill[fill=c07] (2.5,2.5) rectangle (3.5,3.5);   \node[font=\huge] at (3,3) {.67};
    \fill[fill=c08] (3.5,2.5) rectangle (4.5,3.5);   \node[font=\huge, text=white] at (4,3) {.77};

    \fill[fill=c06] (-.5,1.5) rectangle (.5,2.5);  \node[font=\huge] at (0,2) {.44};
    \fill[fill=c08] (.5,1.5) rectangle (1.5,2.5);   \node[font=\huge, text=white] at (1,2) {.7};
    \fill[fill=c10] (1.5,1.5) rectangle (2.5,2.5);   \node[font=\huge, text=white] at (2,2) {1};
    \fill[fill=c07] (2.5,1.5) rectangle (3.5,2.5);   \node[font=\huge] at (3,2) {.63};
    \fill[fill=c07] (3.5,1.5) rectangle (4.5,2.5);   \node[font=\huge] at (4,2) {.69};

    \fill[fill=c06] (-.5,.5) rectangle (.5,1.5);  \node[font=\huge] at (0,1) {.23};
    \fill[fill=c07] (.5,.5) rectangle (1.5,1.5);   \node[font=\huge] at (1,1) {.67};
    \fill[fill=c07] (1.5,.5) rectangle (2.5,1.5);   \node[font=\huge] at (2,1) {.63};
    \fill[fill=c10] (2.5,.5) rectangle (3.5,1.5);   \node[font=\huge, text=white] at (3,1) {1};
    \fill[fill=c08] (3.5,.5) rectangle (4.5,1.5);   \node[font=\huge, text=white] at (4,1) {.73};

    \fill[fill=c06] (-.5,-.5) rectangle (.5,.5);  \node[font=\huge] at (0,0) {.27};
    \fill[fill=c08] (.5,-.5) rectangle (1.5,.5);   \node[font=\huge, text=white] at (1,0) {.77};
    \fill[fill=c07] (1.5,-.5) rectangle (2.5,.5);   \node[font=\huge] at (2,0) {.69};
    \fill[fill=c08] (2.5,-.5) rectangle (3.5,.5);   \node[font=\huge, text=white] at (3,0) {.73};
    \fill[fill=c10] (3.5,-.5) rectangle (4.5,.5);   \node[font=\huge, text=white] at (4,0) {1};

    \node[font=\huge, rotate=45, anchor=east] at (0,-.7) {LCID4 RX};
    \node[font=\huge, rotate=45, anchor=east] at (1,-.7) {LCID4 Tput};
    \node[font=\huge, rotate=45, anchor=east] at (2,-.7) {MAC RX};
    \node[font=\huge, rotate=45, anchor=east] at (3,-.7) {SNR};
    \node[font=\huge, rotate=45, anchor=east] at (4,-.7) {MAC Tput};

    \node[font=\huge, anchor=east] at (-.7,4) {LCID4 RX};
    \node[font=\huge, anchor=east] at (-.7,3) {LCID4 Tput};
    \node[font=\huge, anchor=east] at (-.7,2) {MAC RX};
    \node[font=\huge, anchor=east] at (-.7,1) {SNR};
    \node[font=\huge, anchor=east] at (-.7,0) {MAC Tput};

    \draw (-.5,-.5) rectangle (4.5,4.5);
\end{tikzpicture}}\
     \setlength\abovecaptionskip{1.0cm}
    \setlength\belowcaptionskip{-1.0cm}
    \caption{OAI cluster.}
    \label{fig:corr_oai}
    \vspace{1.0cm}
\end{subfigure}
    \definecolor{c10}{RGB}{5,48,97}
\definecolor{c09}{RGB}{33,102,172}
\definecolor{c08}{RGB}{67,147,195}
\definecolor{c07}{RGB}{146,197,222}
\definecolor{c06}{RGB}{209,229,240}
\definecolor{c05}{RGB}{247,247,247}
\definecolor{c04}{RGB}{253,219,199}
\definecolor{c03}{RGB}{244,165,130}
\definecolor{c02}{RGB}{214,96,77}
\definecolor{c01}{RGB}{178,24,43}
\definecolor{c00}{RGB}{103,0,31}
\begin{tikzpicture}
    \pgfplotsset{
        colormap={RdBu}{
            rgb255=(103,0,31)
            rgb255=(178,24,43)
            rgb255=(214,96,77)
            rgb255=(244,165,130)
            rgb255=(253,219,199)
            rgb255=(247,247,247)
            rgb255=(209,229,240)
            rgb255=(146,197,222)
            rgb255=(67,147,195)
            rgb255=(33,102,172)
            rgb255=(5,48,97)
        }
    }
    \begin{axis}[
        hide axis,
        scale only axis,
        height=0pt,
        width=0pt,
        colormap name=RdBu,
        colorbar horizontal,
        point meta min=-1,
        point meta max=1,
colorbar style={
    width=0.7\columnwidth,   
    height=0.15cm,
    xtick={-1,-0.5,0,0.5,1},
    xticklabel style={font=\scriptsize},
    title={\scriptsize Correlation Coefficient},
    title style={yshift=-0.15cm},
    at={(.5,0)},  
    anchor=center,
},
    ]
        \addplot [draw=none] coordinates {(0,0)};
    \end{axis}
\end{tikzpicture}
    \caption{Clustered correlation matrices for Data Lake and OAI KPMs, revealing redundancy structure used to select a compact input set for the switching policy.}
    \label{fig:correlation_matrices}
\end{figure}

While the preceding analysis identifies \glspl{kpm} that are individually sensitive to channel estimation quality, several of these metrics may carry redundant information. This redundancy arises because metrics are tightly coupled through link adaptation (e.g., MCS index and TB size move in lockstep as the scheduler adapts to channel quality). To distill the \gls{kpm} set into a compact and non-redundant input for the \switch dApp, we compute pairwise Pearson correlation coefficients across all candidate \glspl{kpm} and reorder the resulting matrices using hierarchical clustering, which groups highly correlated \glspl{kpm} into visually identifiable block-like structures. \gls{phy} throughput from Aerial is excluded from the correlation analysis due to its cumulative computation, unlike the other per-slot metrics from Aerial.

Figure~\ref{fig:corr_datalake} shows the correlation matrix for the Aerial Data Lake \glspl{kpm}. We apply a conservative correlation threshold of 0.8, chosen to retain \glspl{kpm} that, despite moderate correlation, may carry complementary information under specific channel conditions. We observe only one redundant cluster, visible as the darker blue shaded block containing code rate, \gls{sinr}, \gls{qam} order, \gls{mcs} index, \gls{tb} size, and the number of \glspl{cb}, with pairwise correlation coefficients ranging from 0.81 to 1. Link adaptation causes these parameters to change together as channel quality varies.
From this cluster, we keep \gls{mcs} index as the representative, as it directly reflects the scheduling decision and subsumes the effect of the remaining parameters. The remaining \glspl{kpm} (\gls{pdu} length, \gls{ndi}, and \gls{rsrp}) fall below the threshold with respect to each other and are preserved as independent inputs to the switching logic. Notably, \gls{rsrp} exhibits a consistent negative correlation with the main block (from $-0.58$ to $-0.71$). Since \gls{rsrp} in Aerial is computed from the mean squared magnitude of the channel estimates, additive noise inflates the measured power ($\text{RSRP} \approx \text{RSRP}_{\text{clean}} + \sigma^2$). As a result, \gls{rsrp} increases as channel estimation quality degrades, while the remaining \glspl{kpm} decline.


Figure~\ref{fig:corr_oai} shows the corresponding matrix for the \gls{oai} \glspl{kpm}. No pairwise correlation exceeds the 0.8 threshold, with the strongest association being \gls{mac} throughput and LCID4 throughput (0.77). Consequently, all \gls{oai}-derived \glspl{kpm} are retained. The generally moderate correlations suggest that these \glspl{kpm} capture complementary aspects of system performance, for instance, LCID4 RX exhibits the weakest correlations overall (0.23--0.46), reflecting user-plane delivery rather than \gls{phy}-layer link adaptation, and providing the \switch dApp with a cross-layer view of system health.

Based on this analysis, the final set of \glspl{kpm} selected for the \switch switching logic comprises: \gls{phy} throughput, \gls{mcs} index, \gls{pdu} length, \gls{ndi}, and \gls{rsrp} from Aerial Data Lake; and \gls{snr}, \gls{mac}-layer throughput, LCID4 throughput, \gls{mac} received bytes, and LCID4 received bytes from \gls{oai}. Together, these \glspl{kpm} span multiple protocol layers, from \gls{phy}-layer channel quality metrics to \gls{mac}-layer scheduling and throughput indicators, giving \switch a multi-layer perspective on channel conditions that no single metric could provide alone.

\vspace{-.4cm}
\section{Case Study: \gls{ul} Channel Estimation}
\label{sec:casestudy}

This section instantiates the \switch framework for \gls{ul} channel estimation, detailing the domain-specific aspects: the two experts (\gls{ai}-based and \gls{mmse}), the background on channel estimation, and the specific switching policy.

\vspace{-.3cm}
\subsection{Background: Channel Estimation}
\label{sec:casestudy_bg}

\gls{ul} channel estimation at the \gls{gnb} is performed using \gls{dmrs} embedded within the \gls{pusch}. The \gls{dmrs} patterns and sequences are predefined by the standard and configured by the \gls{gnb}, with the corresponding parameters conveyed to the \gls{ue} through Downlink Control Information (DCI). Upon reception, the \gls{gnb} correlates the known \gls{dmrs} sequences with the received signal at the \gls{dmrs} locations to estimate the channel response, thereby characterizing the effects of propagation, fading, and other impairments.

\begin{figure}[htbp]
  \centering
      \includegraphics[width=1\columnwidth]{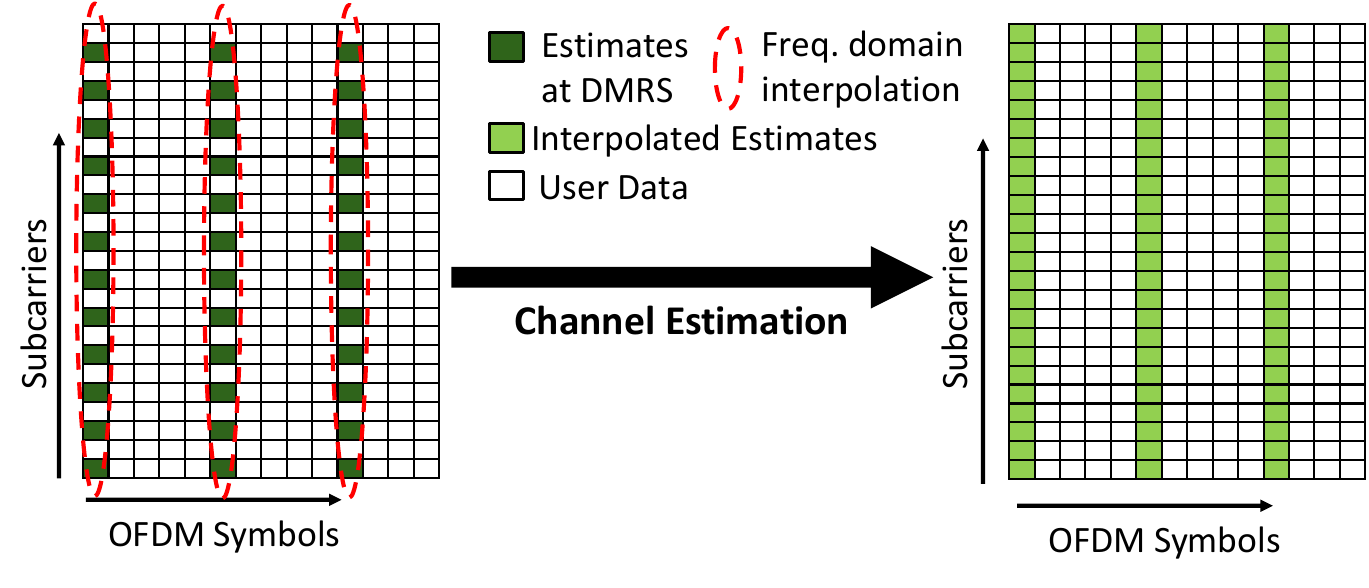}
\setlength\abovecaptionskip{-.5cm}
  \caption{DMRS-Based Channel Estimation and Frequency-Domain Interpolation in a 5G NR UL Slot.}
  \label{fig:dmrs}
\end{figure}

These initial channel estimates are then interpolated in the frequency domain across all subcarriers to obtain channel estimates over the entire frequency band, as shown in Figure~\ref{fig:dmrs}.
The left image in Figure~\ref{fig:dmrs} shows a representative \gls{ofdm} \gls{ul} slot, where user data resource elements are shown in white and the channel estimates computed at the \gls{dmrs} resource elements are shown in dark-green. We use \gls{dmrs} with a Type 1 configuration with interleaved frequency-domain placement, occupying OFDM symbols 0, 5, and 10.
The right-hand image in Figure~\ref{fig:dmrs} illustrates the frequency-domain interpolated channel estimates in light-green, where the initial estimates at the \gls{dmrs} resource elements are extended across all subcarriers within the corresponding \gls{ofdm} symbols, thereby reconstructing the channel over the entire bandwidth.

The interpolation and smoothing strategy depends on the specific estimator used. Some estimators also perform time-domain interpolation. In the current implementation of \switch on NVIDIA Aerial, however, the channel estimator performs only frequency-domain interpolation, while time-domain interpolation across \gls{ofdm} symbols is handled by the subsequent equalizer.

The \gls{mmse} estimator leverages prior knowledge of channel and noise statistics but relies on assumptions (accurate covariance matrices, quasi-stationary statistics) that limit its robustness under time-selective channels, strong multipath, or statistical mismatch. Various neural-network estimators have shown promising performance under specific conditions, but no single estimator dominates across all scenarios~\cite{feriani2023cebed}. This motivates the \switch approach: rather than committing to one estimator, the framework maintains both and selects per-slot.

\vspace{-.35cm}
\subsection{Expert Bank: \gls{ai}-Based and \gls{mmse} Estimators} 
\label{sec:expertbank}

The \gls{pusch} CUDA graph (Figure~\ref{fig:prototype}) is 
extended to execute both the \gls{mmse} (node~2c) and \gls{ai}-based 
(node~2d) estimators in parallel, with the switch kernel 
(node~2e) selecting the output. For this instantiation, 
\texttt{mode}=0 selects \gls{ai}-based estimates and \texttt{mode}=1 
selects \gls{mmse}. Separate memory is allocated for 
$\hat{\mathbf{H}}_{\mathrm{MMSE}}$ and 
$\hat{\mathbf{H}}_{\mathrm{AI}}$, with downstream modules 
reading from the \gls{ai}-designated buffer.

The \gls{mmse} estimator (Expert~A) is the conventional channel 
estimator native to the NVIDIA Aerial \gls{pusch} pipeline, 
performing frequency-domain interpolation based on power-delay 
profile approximation~\cite{hung2010pilot}, as described in 
Section~\ref{sec:casestudy_bg}.

The \gls{ai}-based estimator (Expert~B) employs a deep neural network 
with convolutional layers and residual connections, following 
architectures that have shown strong performance for OFDM channel 
estimation~\cite{li2020reesnet}, taking 
\gls{ls} channel estimates at \gls{dmrs} locations as input and 
producing frequency-interpolated estimates across the full 
bandwidth. The model is integrated via the Channel-Estimate 
Factory on the Aerial platform using the TensorRT inference 
engine, and is tested in the open-source cuBB 25-3~\cite{nvidia25}. 
The current NVIDIA Aerial integration supports \gls{ai}-based estimation 
for a single \gls{ue}; extending to multi-\gls{ue} scheduling is 
left to future work. We will release the \switch extensions, 
including the switch kernel and dApp integration, as open source components. 
The \gls{ai}-based estimator's inference time and \gls{gpu} utilization are 
reported in Section~\ref{sec:perf}.

\vspace{-0.3cm}
\subsection{Switching Policy: Decision Tree}
\label{sec:decision}
The switching policy is implemented as a lightweight decision 
tree classifier in the \switch dApp. It takes as input the 
\glspl{kpm} selected via the methodology in 
Section~\ref{sec:methodology} and outputs the binary 
\texttt{mode} variable.

\textbf{Training data and labeling.} The decision tree is trained via supervised learning on \gls{ota} data, where each slot is labeled \texttt{mode}=0 under interference conditions and \texttt{mode}=1 otherwise. The goal is to identify when the additional performance from \gls{ai} justifies its higher compute and energy cost (Section~\ref{sec:perf}). We use interference presence as the supervisory signal because \gls{ota} profiling shows that \gls{ai} provides its largest gains under interference, while \gls{mmse} remains near-optimal in clean conditions.


\textbf{Model configuration.} The tree has a maximum depth of 2
with Gini impurity splitting across 10 \glspl{kpm}. The most
informative features are \gls{mac}-layer throughput (94.27\%), \gls{mac}
received bytes (4.30\%), and LCID4 throughput (1.42\%). The
shallow depth ensures sub-microsecond inference while avoiding
overfitting.

\textbf{Performance.} As shown in Table~\ref{tab:dt_performance}, the decision tree achieves 
99.66\% accuracy, 97.56\% precision, 99.60\% specificity, and a 98.77\% 
F1 score on a held-out test set (80/20 split).

\begin{table}[h]
\centering
\caption{Decision tree classification performance.}
\label{tab:dt_performance}
\begin{tabular}{cccc}
\toprule
\textbf{Accuracy} & \textbf{Precision} & \textbf{Specificity} & \textbf{F1 Score} \\
\midrule
99.66\% & 97.56\% & 99.60\% & 98.77\% \\
\bottomrule
\end{tabular}
\end{table}

\vspace{-.4cm}

\section{Experimental Evaluation}
\label{sec:eval}

\begin{figure}[htbp]
  \centering
  \begin{subfigure}[b]{\columnwidth}
    \centering
    \includegraphics[width=\columnwidth]{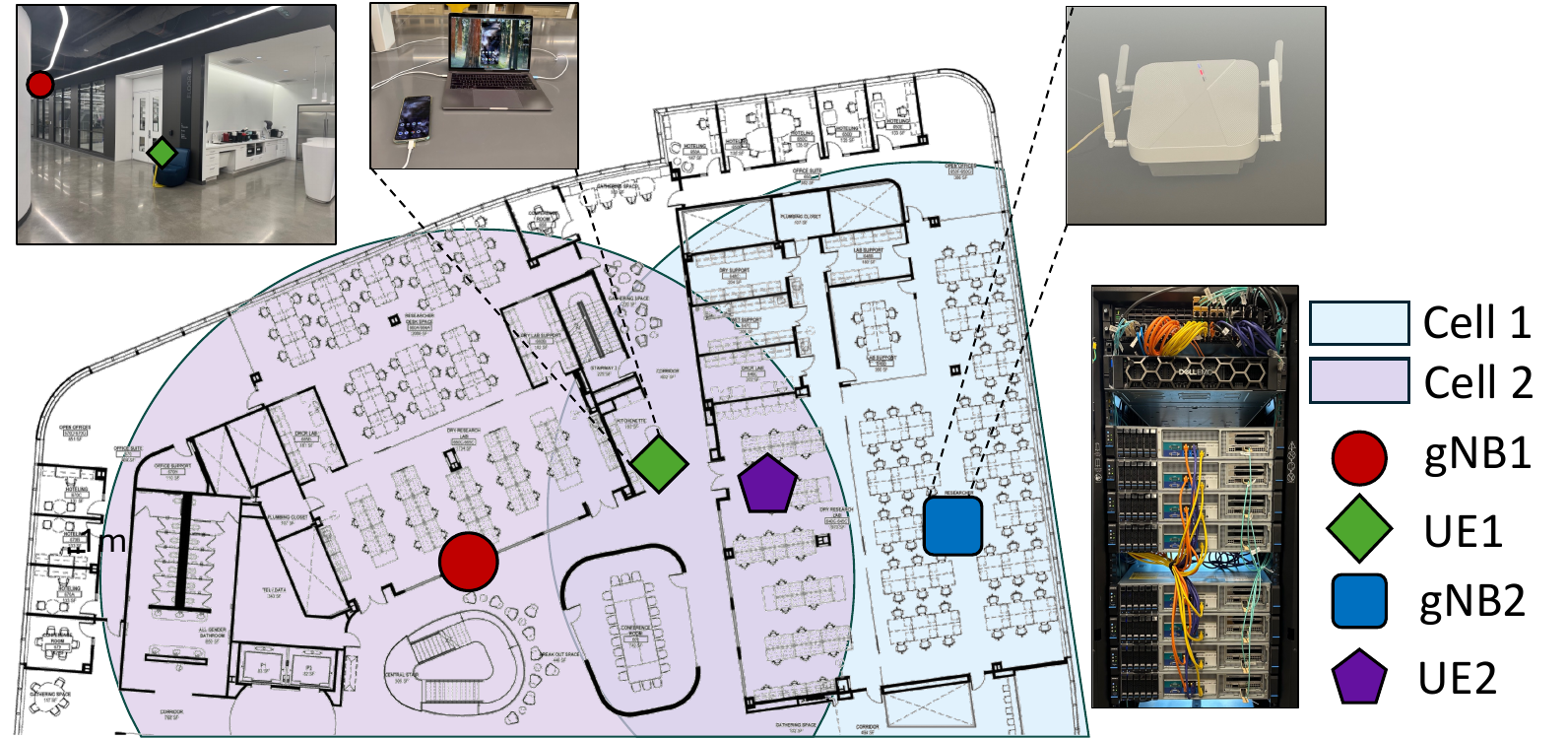}
    \caption{Floor plan showing two cell sites. Insets: UE (top center), RF environment for gNB1--UE1 (top left), O-RU (top right), server (bottom right).}
    \label{fig:exp1}
  \end{subfigure}
  \\[4pt]
  \begin{subfigure}[b]{1\columnwidth}
    \centering
    \includegraphics[width=.65\linewidth]{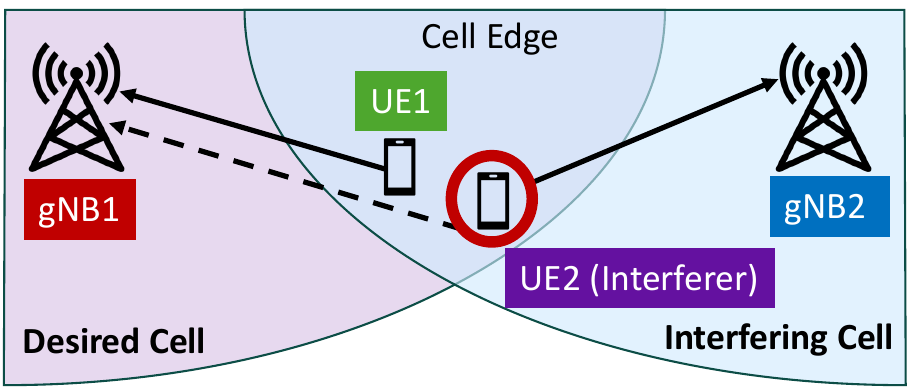}
    \caption{Introduction of interference via UL transmissions between UE2 and gNB2.}
    \label{fig:interf1}
  \end{subfigure}
  \caption{Experimental setup.}
  \label{fig:exp_setup}
\end{figure}

\textbf{Aerial Data Lake.} To collect the \glspl{kpm} that drive the switching policy, we 
use NVIDIA Aerial Data Lake, a telemetry framework within the NVIDIA Aerial CUDA-accelerated \gls{ran} that captures \gls{ul} I/Q samples from the O-RAN 7.2x fronthaul interface along with FAPI metadata exchanged between the L1 and L2 layers.

\textbf{OTA data collection.} All experiments are conducted on 
the X5G platform~\cite{villa2025tmc}, an O-RAN--compliant, 
multi-vendor private 5G network built on NVIDIA 
Aerial~\cite{arc-ota}, covering an entire floor of the EXP 
building at Northeastern University. Each node features \gls{oai} 
for upper layers and the NVIDIA Aerial \gls{sdk} for the DU-Low, 
running on GH200 \gls{gpu} systems.

The \gls{ota} experiments are conducted in an indoor open space with a 
dynamic environment (Figure~\ref{fig:exp1}). All measurements are performed 
under \gls{los} conditions between the \gls{gnb}1--\gls{ue}1 
pair, with \gls{ue}1 placed on a stand and \gls{gnb}1 mounted on 
a wall (Figure~\ref{fig:exp1}). To evaluate performance under 
different conditions, we introduce frequency-selective in-band 
\gls{ul} interference from a neighboring cell~\cite{interforan}: 
\gls{ue}2 transmits to \gls{gnb}2 in the same frequency band, 
with the interference level controlled by the \gls{mac} scheduler via 
\gls{prb} allocation (Figure~\ref{fig:interf1}). We refer to 
conditions as \emph{good} (no interference) and \emph{poor} 
(interference present).

\vspace{-.3cm}
\subsection{Mechanism-Level Results: Temporal Analysis}
\label{sec:latency}

\begin{figure}[htbp]
    \centering
    \setlength\fwidth{1\linewidth}
    \setlength\fheight{.6\linewidth}
            \begin{tikzpicture}
\definecolor{aicolor}{RGB}{32,178,170}
\definecolor{mmsecolor}{RGB}{144,238,200}
\definecolor{aiestcolor}{RGB}{70,130,180}
\definecolor{mmseestcolor}{RGB}{255,165,80}
\definecolor{dtcolor}{RGB}{186,85,211}
\begin{axis}[
    width=1\fwidth,
    height=1\fheight,
    ybar,
    bar width=10pt,
    enlarge x limits=0.2,
    symbolic x coords={Average, Median, Std Dev},
    xtick=data,
    x tick label style={font=\small},
    ylabel={Time [$\mu$s]},
    ylabel style={font=\small},
    point meta=rawy,
    title style={font=\normalsize},
    legend style={
        at={(.51,1.02)},
        anchor=south,
            legend columns=5,
        font=\footnotesize
    },
    ymode=log,
    log origin=infty,
    ymin=0.1, ymax=1000,
    yminorticks=false,
    nodes near coords,
    nodes near coords style={font=\footnotesize, /pgf/number format/fixed, /pgf/number format/precision=2, /pgf/number format/1000 sep={},},
    every node near coord/.append style={anchor=south},    
    grid=major,
    grid style={dashed, gray!30},
    tick label style={font=\small},
]
\addplot[fill=aicolor, 
    draw=aicolor!40!black,
    postaction={pattern=north east lines, pattern color=black}]
    coordinates {
        (Average, 3.36)+- (0.0097, 0.0097) 
        (Median,  3.36)
        (Std Dev, .41)
    };
    
\addplot[fill=mmsecolor, 
    draw=mmsecolor!40!black,
    postaction={pattern=horizontal lines, pattern color=black}]
    coordinates {
        (Average, 4.89)
        (Median,  4.83)
        (Std Dev, .47)
    };
\addplot[fill=aiestcolor, 
    draw=aiestcolor!40!black,
    postaction={pattern=north west lines, pattern color=black}]
    coordinates {
(Average, 432.329) +- (0.052, 0.052)
(Median, 432.373) +- (0.000, 0.000)
(Std Dev, 0.842) +- (0.010, 0.009)
    };
\addplot[fill=mmseestcolor, 
    draw=mmseestcolor!40!black,
    postaction={pattern=dots, pattern color=black}]
    coordinates {
(Average,  5.0406)  +-  (0.0056, 0.0056)   
(Median,   4.6080)  +-  (0.8640, 6.6880)  
(Std Dev,  0.8330)  +-  (0.0040, 0.0043)  
    };
\addplot[fill=dtcolor, 
    draw=dtcolor!40!black,
    postaction={pattern=crosshatch, pattern color=black}]
    coordinates {
        (Average, 0.41) +- (0.0097, 0.0097)
        (Median,  0.38) +- (0.0320, 0.0000) 
        (Std Dev, 0.12) +- (0.0063, 0.0074) 
    };
\legend{Switch (AI), Switch (MMSE), AI, MMSE, Decision Tree}
\end{axis}
\end{tikzpicture}
        \captionsetup{aboveskip=0pt, belowskip=.1pt} 
    \caption{Runtime statistics for all components of the switching mechanism.}
    \label{fig:timing}
\end{figure}
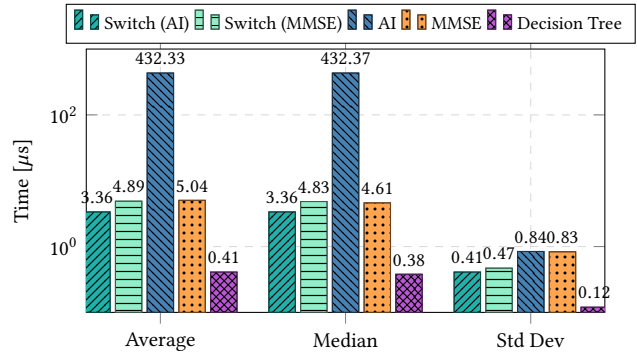

\textbf{Switch Kernel.} Figure~\ref{fig:timing} summarizes the 
runtime statistics for the \gls{ai} expert (dark-green) and the 
\gls{mmse} estimator (light-green), profiled using NVIDIA Nsight 
Systems. The switch kernel executes in an average of 3.36~$\mu$s 
when selecting \gls{ai} and 4.89~$\mu$s when selecting \gls{mmse}. The 
higher latency in \gls{mmse} mode reflects the coalesced memory copy 
discussed in Section~\ref{sec:switch_kernel}, which is absent in 
\gls{ai} mode. Similar trends are observed across the median, and 
standard deviations remain comparable for both modes, indicating 
consistent execution.

\textbf{Decision Tree.} The inference time of the decision tree (purple bars in Figure~\ref{fig:timing}), described in Section~\ref{sec:decision}, running on the \switch dApp averages 0.41~$\mu$s on the GH200 (median 0.38~$\mu$s, std.\ dev.\ 0.12~$\mu$s), achieving an accuracy above 99\%. The model is executed at a configurable periodicity, enabling timely adaptation to channel dynamics without introducing measurable processing overhead.

\textbf{End-to-end Control Loop Latency.} The end-to-end control loop latency accounts for the full data path from the \gls{gpu}-accelerated \gls{pusch} pipeline to the dApp, including the decision tree inference and the switch kernel execution. Prior to any model execution, the dApp incurs approximately 135~$\mu$s of framework overhead, comprising several tightly optimized data-movement and messaging steps, such as copying data into shared memory buffers and sending pointers and metadata to the dApp via the ZeroMQ interface. The decision tree classifier adds 0.41~$\mu$s on average, and the switch kernel itself executes in approximately 4.5~$\mu$s, yielding a total end-to-end control loop latency of approximately 140~$\mu$s. Note that this latency assumes the switching decision is applied at the beginning of the current slot; if the decision arrives after the slot boundary, it takes effect in the following slot.

\textbf{Data Integrity.} Since \gls{phy} throughput in Aerial is computed from successfully decoded transport blocks based on \gls{tb} \gls{crc} checks, any corruption introduced by switching would therefore appear as reduced throughput. The stable throughput observed during switching confirms that slot-boundary switching does not corrupt in-flight data.

\vspace{-.3cm}
\subsection{Case-Study Results: Performance Comparison}
\label{sec:perf}


\textbf{Expert Inference Times.} We profile the execution time of the \gls{ai}-based estimator (blue) and the \gls{mmse} filtering kernel (orange) in Figure~\ref{fig:timing}. The \gls{ai}-based estimator inference time is profiled using TensorRT, while the \gls{mmse} latency is measured using Nsight Systems by profiling the dispatch kernel, which internally invokes the kernel that performs the actual \gls{mmse} estimation. The \gls{mmse} CUDA kernel executes in approximately 5.04~$\mu$s, while the \gls{ai}-based estimator requires approximately 432~$\mu$s, making it roughly 85 times slower. This overhead, along with the \gls{gpu} utilization discussed later in this subsection, is the primary motivation for the switching mechanism proposed in this work. Although the \gls{ai}-based estimator delivers consistent throughput gains under the tested conditions, its substantially higher compute cost makes unconditional execution impractical, particularly in multi-\gls{ue} scenarios where per-slot latency budgets are shared across users. The dApp therefore dynamically activates the \gls{ai}-based estimator only when the switching policy determines that conditions warrant it.

\begin{figure}[htbp]
    \centering
    \setlength\fwidth{1\linewidth}
    \setlength\fheight{1\linewidth}
            \input{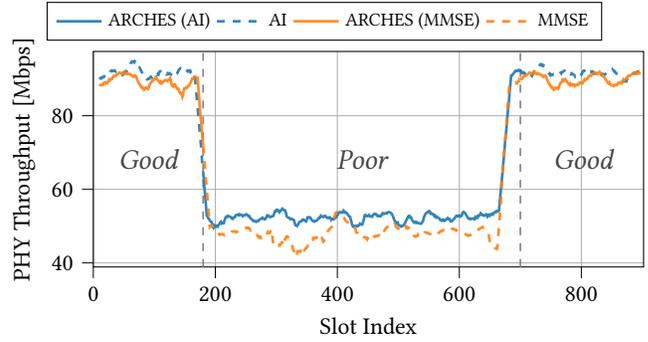}
        \captionsetup{aboveskip=-8pt, belowskip=.1pt} 
    \caption{PHY throughput over time as channel conditions transition from good to poor and back. Dashed lines: throughput under continuous execution of each estimator; solid lines: throughput under \switch.}
    \label{fig:time_series}
\end{figure}

\textbf{Time Series of \gls{phy} Throughput.} Figure~\ref{fig:time_series} shows the \gls{phy} throughput over time (indexed by slot) for an \gls{ota} experiment in which channel conditions transition from good to poor and back. Dashed lines show the throughput each estimator would achieve if run continuously; solid lines show the throughput under ARCHES. 

In good conditions, \switch selects \gls{mmse}: although AI consistently improves throughput, the gain is too small to justify its substantially higher compute and energy cost, as quantified under \textit{\gls{gpu} and CPU Resources} below. In poor conditions, \switch switches to the \gls{ai}-based estimator, where the larger throughput advantage outweighs the additional resource consumption.

\begin{figure}[htbp]
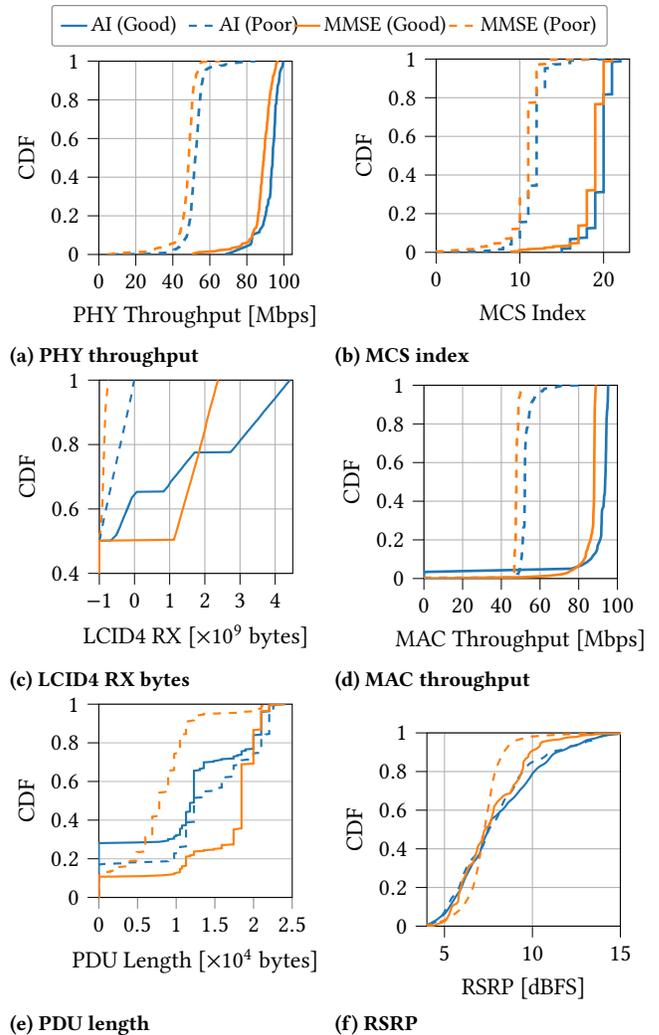

    \centering
    \begin{tikzpicture}
        \definecolor{myblue}{RGB}{31,119,180}
        \definecolor{myorange}{RGB}{255,127,14}
        \node[draw=gray, rounded corners=1pt, inner sep=3pt, fill=white] {
            \begin{tikzpicture}[baseline=0pt]
                \draw[thick, myblue] (0,0) -- (0.35,0) node[right, black, font=\small, inner sep=1pt] {AI (Good)};
                \draw[thick, myblue, dashed] (1.7,0) -- (2.05,0) node[right, black, font=\small, inner sep=1pt] {AI (Poor)};
                \draw[thick, myorange] (3.1,0) -- (3.45,0) node[right, black, font=\small, inner sep=1pt] {MMSE (Good)};
                \draw[thick, myorange, dashed] (5.16,0) -- (5.53,0) node[right, black, font=\small, inner sep=1pt] {MMSE (Poor)};
            \end{tikzpicture}
        };
    \end{tikzpicture}
    \\[-2pt]
    \begin{subfigure}[b]{0.49\linewidth}
        \centering
        \input{figures_tex/cdf_thru}
            \captionsetup{aboveskip=-8pt, belowskip=-3pt}
        \caption{PHY throughput}
        \label{fig:cdf_thru}
    \end{subfigure}
    \hfill
    \begin{subfigure}[b]{0.49\linewidth}
        \centering
        \input{figures_tex/cdf_mcs_datalake}
        \captionsetup{aboveskip=6pt, belowskip=-3pt}
        \caption{MCS index}
        \label{fig:cdf_mcs_datalake}
    \end{subfigure}

    \begin{subfigure}[b]{0.49\linewidth}
        \centering
        \input{figures_tex/cdf_lcid4}
        \captionsetup{aboveskip=4pt, belowskip=-3pt}
        \caption{LCID4 RX bytes}
        \label{fig:cdf_lcid4}
    \end{subfigure}
    \hfill
    \begin{subfigure}[b]{0.49\linewidth}
        \centering
        \input{figures_tex/cdf_macthru}
        \captionsetup{aboveskip=-8pt, belowskip=-3pt}
        \caption{MAC throughput}
        \label{fig:cdf_macthru}
    \end{subfigure}

        \begin{subfigure}[b]{0.49\linewidth}
        \centering
        \input{figures_tex/cdf_pdulen}
        \captionsetup{aboveskip=0.5pt, belowskip=-3pt}
        \caption{PDU length}
        \label{fig:cdf_pdulen}
    \end{subfigure}
    \hfill
    \begin{subfigure}[b]{0.49\linewidth}
        \centering
        \input{figures_tex/cdf_rsrp1}
        \captionsetup{aboveskip=2pt, belowskip=-3pt}
        \caption{RSRP}
        \label{fig:cdf_rsrp}
    \end{subfigure}
    \caption{CDF comparison of AI (blue) and MMSE (orange) estimators under good (solid) and poor (dashed) channel conditions.}
    \label{fig:cdf_comparison_KPM}
\end{figure}

\textbf{KPMs.} Figure~\ref{fig:cdf_comparison_KPM} presents CDFs of six \glspl{kpm} used in the switching logic. Under good conditions, the \gls{ai}-based estimator achieves a median \gls{phy} throughput of 93.97~Mbps versus 89.23~Mbps for \gls{mmse} (5.32\% gain) and a median \gls{mac} throughput of 93.67~Mbps versus 87.99~Mbps (6.45\% gain). Under poor conditions, the \gls{ai} advantage persists: median \gls{phy} throughput of 52.14 versus 48.64~Mbps (7.23\%) and \gls{mac} throughput of 52.18 versus 47.77~Mbps (9.23\%). For \gls{mcs} index (Figure~\ref{fig:cdf_mcs_datalake}), the \gls{ai}-based estimator yields a median of 20 versus 19 under good conditions and 12 versus 11 under poor conditions.

For PDU length (Figure~\ref{fig:cdf_pdulen}), both estimators produce larger PDUs under good conditions, with \gls{ai} shifting the distribution toward higher values. LCID4 RX bytes (Figure~\ref{fig:cdf_lcid4}) follow a similar trend, confirming that \gls{phy}-layer gains propagate to user-plane delivery. RSRP (Figure~\ref{fig:cdf_rsrp}) remains comparable across estimators, consistent with its dependence on received signal power rather than estimation quality.

\begin{figure}[htbp]
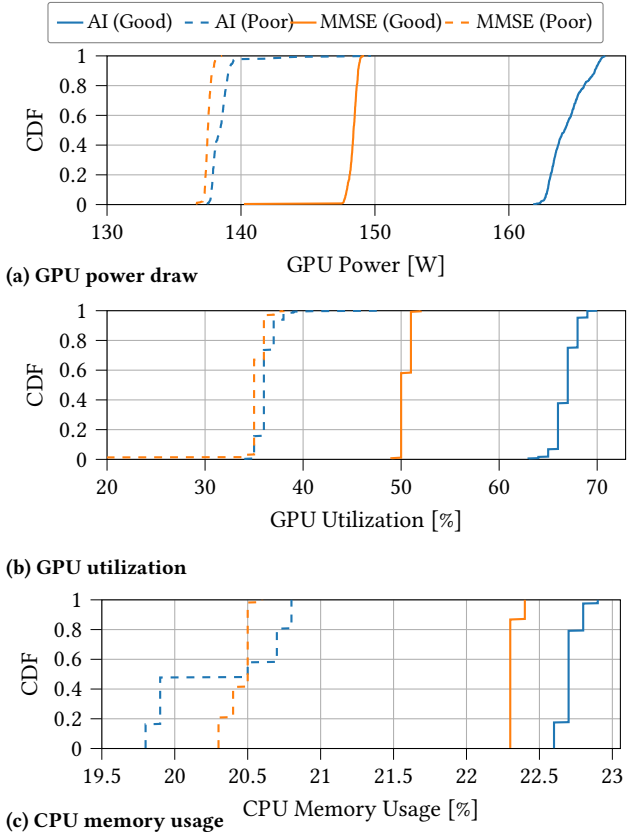

    \centering
    \setlength\fheight{0.3\linewidth}
    \begin{tikzpicture}
        \definecolor{myblue}{RGB}{31,119,180}
        \definecolor{myorange}{RGB}{255,127,14}
        \node[draw=gray, rounded corners=1pt, inner sep=3pt, fill=white] {
            \begin{tikzpicture}[baseline=0pt]
                \draw[thick, myblue] (0,0) -- (0.35,0) node[right, black, font=\small, inner sep=1pt] {AI (Good)};
                \draw[thick, myblue, dashed] (1.7,0) -- (2.05,0) node[right, black, font=\small, inner sep=1pt] {AI (Poor)};
                \draw[thick, myorange] (3.1,0) -- (3.45,0) node[right, black, font=\small, inner sep=1pt] {MMSE (Good)};
                \draw[thick, myorange, dashed] (5.16,0) -- (5.53,0) node[right, black, font=\small, inner sep=1pt] {MMSE (Poor)};
            \end{tikzpicture}
        };
    \end{tikzpicture}
    \\[-2pt]
    \begin{subfigure}[b]{\linewidth}
        \centering
        \input{figures_tex/cdf_gpu_power}
            \captionsetup{aboveskip=-8pt, belowskip=1pt}
        \caption{GPU power draw}
        \label{fig:cdf_gpu_power}
    \end{subfigure}
    \begin{subfigure}[b]{\linewidth}
        \centering
        \input{figures_tex/cdf_gpu_util}
        \caption{GPU utilization}
        \label{fig:cdf_gpu_util}
    \end{subfigure}
        \begin{subfigure}[b]{\linewidth}
        \centering
        \input{figures_tex/cdf_cpu_mem}
            \captionsetup{aboveskip=-8pt, belowskip=1pt}
        \caption{CPU memory usage}
        \label{fig:cdf_cpu_memory}
    \end{subfigure}
       \captionsetup{aboveskip=4pt, belowskip=1pt}
    \caption{CDF comparison of GPU power draw, GPU utilization, and CPU memory usage for AI (blue) and MMSE (orange) estimators under good (solid) and poor (dashed) channel conditions.}
    \label{fig:cdf_comparison_gpu}
\end{figure}

\textbf{\gls{gpu} and CPU Resources.} Figure~\ref{fig:cdf_comparison_gpu} presents the CDFs of \gls{gpu} power, utilization, and CPU memory. Measurements are collected with each estimator running independently, quantifying the per-expert resource profile that informs the switching decision. Under good conditions, \gls{mmse} shows a clear advantage: median \gls{gpu} power decreases from 164.2~W (\gls{ai}) to 148.4~W, saving 15.8~W (9.6\%), while \gls{gpu} utilization drops from 67\% to 50\% (17 percentage points). Under poor conditions, reduced scheduling grants lower overall \gls{gpu} load for both estimators, shrinking the power gap to \(\sim\)1~W and converging utilization to 36\% versus 35\%. CPU memory remains comparable across all cases, confirming negligible host-side overhead. The resource gap is most pronounced under good conditions, thus reinforcing the case for defaulting to MMSE when conditions are favorable.

Concurrent execution is used here to validate seamless slot-boundary switching, deterministic latency, and fail-safe operation. In practical deployment, the selected-only mode (Section~\ref{sec:expert_bank}) would directly realize these savings at the cost of at least a one-slot activation delay. These resource profiles validate the \switch design rationale: activating \gls{ai} only when conditions warrant it improves \glspl{kpm} while avoiding unnecessary power expenditure.

\vspace{-.4cm}
\section{Generalization to Other \gls{ran} Functions}
\label{sec:generalization}

The \switch mechanism is not specific to channel estimation. The design principles in Section~\ref{sec:principles} are function-agnostic; only the experts and telemetry inputs change per use case. For beam management, multiple prediction models (e.g., pedestrian vs.\ vehicular)~\cite{xue2024aiml_beam} could execute, with the policy driven by beam failure rate and mobility indicators. For equalization or \gls{ldpc} decoding, experts with different configurations (e.g., iteration counts or neural-augmented variants~\cite{nachmani2018neural_ldpc}) could be switched based on \gls{bler} and compute-budget metrics. For interference detection and \gls{isac}, prior work on the same platform~\cite{interforan,villa2025cusense} could be extended with multiple detection models selected based on observed interference patterns. In all cases, the validated mechanism-level properties---a 140~$\mu$s control loop, sub-microsecond policy inference, and a 3--5~$\mu$s switch kernel---transfer directly. The switching mechanism remains unchanged; only the experts and telemetry inputs are adapted to the target function.



\vspace{-.4cm}
\section{Related Work}
\label{sec:related}

\textbf{\gls{moe} and adaptive model selection.} The \gls{moe} paradigm~\cite{jacobs1991adaptive,jordan1994hierarchical,shazeer2017outrageously,fedus2022switch} has been widely studied in machine learning, where a gating network routes inputs to specialized sub-models. However, conventional \gls{moe} systems operate in batch or near-real-time settings and do not address the challenges of seamless switching within a live, sub-millisecond signal processing pipeline without data loss. \switch adapts the \gls{moe} concept to the constraints of \gls{gpu}-accelerated 5G \gls{phy} processing.

\textbf{AI-based channel estimation.} Deep learning has emerged as a promising alternative to conventional \gls{mmse} and \gls{ls} estimation in \gls{ofdm} systems. Several works have proposed \gls{cnn} and residual architectures that consider the channel time-frequency responses as 2D images~\cite{soltani2019channelnet,li2020reesnet,feriani2023cebed}, while others have targeted high-mobility scenarios with \gls{rnn}-based interpolators~\cite{gizzini2024rnn} or focused specifically on the 5G NR \gls{pusch} with \gls{ann} and online-trained estimators~\cite{dayi2022ann,weththasinghe2024ml}. A common finding in these works, further highlighted by the CeBed benchmark~\cite{feriani2023cebed}, is that no single estimator dominates across all propagation scenarios, directly motivating condition-aware selection. Notably, ChannelNet~\cite{soltani2019channelnet} itself proposes SNR-based switching between two specialized networks, but remains simulation-only.

\textbf{Real-time AI in \gls{gpu}-accelerated \gls{ran}.} All of the above evaluations are conducted in simulation. Ford \textit{et al.}~\cite{ford2025sim} take a significant step toward real-world validation by demonstrating a ResNet-based \gls{ai} channel estimator on the same NVIDIA testbed platform used in this work, achieving \gls{ota} throughput gains of up to 40\% over \gls{mmse}. Boccuzzi \textit{et al.}~\cite{boccuzzi2025gpu} present a \gls{gpu}-accelerated \gls{ai}-RAN reference design and end-to-end verification methodology on NVIDIA Aerial, while other recent work has demonstrated real-time GPU-accelerated \gls{ai} inference co-located with 5G NR \gls{phy} processing for interference detection and \gls{isac} applications on the same platform~\cite{interforan,villa2025cusense}. However, none of these works incorporates dynamic selection between \gls{ai} and conventional experts; instead, the \gls{ai} model runs continuously, incurring inference overhead regardless of channel conditions.

To our knowledge, this work is the first to present a system for real-time expert switching inside a live GPU-accelerated 5G PHY pipeline, validated end-to-end through OTA experiments with switching decisions driven by a dApp using cross-layer \glspl{kpm} collected from the live system. The contribution of this paper is therefore not a better AI estimator, but the infrastructure required to safely switch between experts in real time.

\vspace{-.3cm}

\section{Conclusions and Future Work}
\label{sec:conclusion}

This paper presents \switch, a framework for real-time expert switching inside \gls{gpu}-accelerated 5G \gls{phy} pipelines. \switch demonstrates that real-time expert switching is practical by enabling multiple \gls{ai} and conventional signal-processing experts to coexist within the pipeline, hot-swapping between them at slot-boundary granularity via a lightweight CUDA kernel, and driving the switching decision through cross-layer telemetry collected by a dApp—all without dropping or corrupting in-flight data. The framework is accompanied by a reusable three-stage process for telemetry selection and policy design.

Validated through a case study on \gls{ul} channel estimation on the X5G platform, \switch achieves an end-to-end control-loop latency of $\sim$140~$\mu$s with sub-microsecond decision inference. The case study demonstrates median \gls{ul} \gls{phy} throughput improvements of 5.32\% and 7.23\% under good and poor conditions, respectively. Critically, under good conditions where the throughput gain from \gls{ai} is marginal ($\sim$5\%), defaulting to \gls{mmse} saves 15.8~W of \gls{gpu} power and 17 percentage points of utilization—quantifying the performance-per-watt tradeoff that motivates condition-aware expert selection as \gls{ai} components proliferate across the \gls{ran}. 

Future work includes extending \switch to support $N > 2$ experts for finer-grained adaptation, scaling the \gls{ai}-based estimator to multi-\gls{ue} scenarios, evaluating additional propagation conditions (\gls{nlos}, high mobility), characterizing the selected-only execution mode under realistic channel dynamics, reporting BLER and HARQ statistics to complement throughput-based evaluation, comparing alternative switching policies such as threshold-based gating, and applying the framework to other real-time \gls{ran} functions.


\vspace{-.3cm}

\balance
\bibliographystyle{IEEEtran}
\bibliography{biblio/sample-base}

\end{document}